
This was created using AMS-TeX:  you should typeset the document
by typing "amstex file".  If you do not have AMS-TeX installed,
you will need to get the AMS-TeX distribution from the ftp site
e-math.ams.org.  You will also need to install the AMS Fonts
package, also available at that site.
\documentstyle{amsppt}

\NoBlackBoxes
\magnification=\magstep1
\hsize=6.5truein
\vsize=8.5truein
\document
\baselineskip=.15truein
\topmatter
\title The $n$-component KP hierarchy and representation theory.
\endtitle
\author
V.G. Kac
and
J.W. van de Leur
\endauthor
\thanks
*Supported in part by the NSF grant DMS-9103792.
\endthanks
\thanks
**Supported by a fellowship of the Royal Netherlands Academy of Arts and
Sciences.
\endthanks
\affil
\endaffil
\address
V.G. Kac, Department of Mathematics, MIT, Cambridge, MA
02139, USA
\endaddress
\address
J.W. van de Leur, Department of Mathematics, University of
Utrecht, Utrecht, The Netherlands
\endaddress
\endtopmatter

\subheading{\S 0. Introduction}

\vskip 10pt
{\bf 0.1.}  The remarkable link between the soliton theory and the group
$GL_{\infty}$ was discovered in the early 1980s by Sato [S] and
developed, making use of the spinor formalism, by Date,
Jimbo, Kashiwara and Miwa [DJKM1,2,3], [JM].   The basic object that they
considered is the KP hierarchy of
partial differential equations, which they study through a sequence
of equivalent formulations that we describe below.  The first
formulation is a deformation (or Lax) equation of a formal
pseudo-differential operator $L = \partial + u_{1}\partial^{-1} +
u_{2}\partial^{-2} + \ldots$, introduced in [S] and [W1]:

$$\frac{\partial L}{\partial x_{n}} = [B_{n},L] ,\ n = 1,2,\ldots .
\tag{0.1.1}$$
Here $u_{i}$ are unknown functions in the indeterminates
$x_{1},x_{2},\ldots$, and $B_{n} = (L^{n})_{+}$ stands for the
differential part of $L^{n}$.  The second formulation is given by
the following zero curvature (or Zakharov-Shabat) equations:

$$\frac{\partial B_{m}}{\partial x_{n}} - \frac{\partial
B_{n}}{\partial x_{m}} = [B_{n},B_{m}],\ m,n = 1,2,\ldots .\tag{0.1.2}$$
These equations are compatibility conditions for the following linear
system

$$Lw(x,z) = zw(x,z) ,\ \frac{\partial}{\partial x_{n}} w(x,z) = B_{n}
w(x,z),\ n = 1,2,\ldots \tag{0.1.3}$$
on the wave function

$$w(x,z) = (1+w_{1}(x)z^{-1} + w_{2}(x)z^{-2} + \ldots
)e^{x_{1}z+x_{2}z^{2}+\ldots}. \tag{0.1.4}$$
Provided that (0.1.2) holds, the system (0.1.3) has a unique solution
of the form (0.1.4) up to multiplication by an element from $1 +
z^{-1}{\Bbb C}[[z^{-1}]]$.  Introduce the wave operator

$$P = 1 + w_{1}(x)\partial^{-1} + w_{2}(x)\partial^{-2} + \ldots ,
\tag{0.1.5}$$
so that $w(x,z) = Pe^{x_{1}z+x_{2}z^{2}+\ldots}$.  Then the existence
of a solution of (0.1.3) is equivalent to the existence of a solution
of the form (0.1.5) of the following Sato equation, which is the
third formulation of the KP hierarchy [S], [W1]:

$$\frac{\partial P}{\partial x_{k}} = -(P \circ \partial \circ
P^{-1})_{-} \circ P, \ k = 1,2,\ldots , \tag{0.1.6}$$
where the formal pseudo-differential operators $P$ and $L$ are
related by

$$L = P \circ \partial \circ P^{-1}. \tag{0.1.7}$$
Let $P^{*} = 1 + (-\partial )^{-1} \circ w_{1} + (-\partial )^{-2}
\circ w_{2} + \ldots$ be the formal adjoint of $P$ and let

$$w^{*}(x,z) = (P^{*})^{-1} e^{-x_{1}z-x_{2}z^{2}-\ldots}$$
be the adjoint wave function.  Then the fourth formulation of the KP
hierarchy is the following bilinear identity
$$\text{Res}_{z=0} w(x,z)w^{*}(x^{\prime},z)dz = 0\ \text{for any}\
x\ \text{and}\ x^{\prime}. \tag{0.1.8}$$
Next, this bilinear identity can be rewritten in terms of Hirota
bilinear operators defined for an arbitary polynomial $Q$ as follows:

$$Q(D)f(x)\cdot g(x) \overset{\text{def}}\to{=} Q(\frac{\partial}{\partial
y})(f(x+y)g(x-y))|_{y=0}. \tag{0.1.9}$$
Towards this end, introduce the famous $\tau$-function $\tau (x)$ by
the formulas:

$$w(x,z) = \Gamma^{+}(z)\tau /\tau , \ w^{*}(x,z) = \Gamma^{-}(z)\tau
/\tau . \tag{0.1.10}$$
Here $\Gamma^{\pm}(z)$ are the vertex operators defined by
$$
\Gamma^{\pm}(z) = e^{\pm (x_{1}z + x_{2}z^{2}+\ldots)} e^{\mp
(z^{-1}{\tilde \partial}/\partial x_{1}+z^{-2}{\tilde
\partial}/\partial x_{2} + \ldots)},
\tag{0.1.11}
$$
where $\frac{\tilde \partial}{\partial x_{j}}$ stands for
$\frac{1}{j} \frac{\partial}{\partial x_{j}}$.  The $\tau$-function
exists and is uniquely determined by the wave function up to a
constant factor.  Substituting the $\tau$-function in the bilinear
identity (0.1.8) we obtain the fifth formulation of the KP hierarchy
as the following system of Hirota bilinear equations:
$$
\sum^{\infty}_{j=0} S_{j}(-2y)S_{j+1}
( \tilde{D} ) e^{\sum^{\infty}_{r=1} y_{r} D_{r}} \tau \cdot \tau = 0.
\tag{0.1.12}
$$
Here $y = (y_{1},y_{2},\ldots )$ are arbitrary parameters and the
elementary Schur polynomials $S_{j}$ are defined by the generating
series

$$\sum_{j \in {\Bbb Z}} S_{j} (x)z^{j} = \exp \sum^{\infty}_{k=1}
x_{k}z^{k}. \tag{0.1.13}$$

The $\tau$-function formulation of the KP hierarchy allows one to
construct easily its $N$-soliton solutions.  For that introduce the
vertex operator [DJKM2,3]:

$$\Gamma (z_{1},z_{2}) = :\Gamma^{+} (z_{1})\Gamma^{-}(z_{2}):
\tag{0.1.14}$$
(where the sign of normal ordering :\ : means that partial
derivatives are always moved to the right), and show using the
bilinear identity (0.1.8) that if $\tau$ is a solution of (0.1.12),
then $(1 + a\Gamma (z_{1},z_{2}))\tau$, where $a,z_{1},z_{2} \in
{\Bbb C}^{\times}$, is a solution as well.   Since $\tau = 1$ is a
solution, the function

$$f_{N} \equiv (1 + a_{1} \Gamma (z^{(1)}_{1},z^{(1)}_{2}))\ldots (1 +
a_{N}\Gamma (z^{(N)}_{1},z^{(N)}_{2})) \cdot 1 \tag{0.1.15}$$
is a solution of (0.1.12) too.  This is the $\tau$-function of the
$N$-soliton solution.

The first application of the KP hierarchy, as well as its name, comes
from the fact that the simplest non-trivial Zakharov-Shabat
equation, namely (0.1.2) with $m = 2$ and $n = 3$, is equivalent to
the Kadomtsev-Petviashvili equation if we let $x_{1} = x,\ x_{2} =
y,\ x_{3} = t,\ u = 2u_{1}$:

$$\frac{3}{4} \frac{\partial^{2}u}{\partial y^{2}} =
\frac{\partial}{\partial x} \left( \frac{\partial u}{\partial t} -
\frac{3}{2} u \frac{\partial u}{\partial x} - \frac{1}{4}
\frac{\partial^{3}u}{\partial x^{3}}\right) . \tag{0.1.16}$$
Recall also that the celebrated KdV and Boussinesq equations are simple
reductions of (0.1.16).  Since the functions $u$ and $\tau$ are
related by
$$u = 2 \frac{\partial^{2}}{\partial x^{2}} \log \tau ,
\tag{0.1.17}$$
the functions $2 \frac{\partial^{2}}{\partial x^{2}} \log f_{N}$ are
solutions of (0.1.16), called the $N$-soliton solutions.

\vskip 10pt
{\bf 0.2.}  The connection of the KP hierarchy to the representation
theory of the group $GL_{\infty}$ is achieved via the spinor
formalism.  Consider the Clifford algebra $C\ell$ on generators
$\psi^{+}_{j}$ and $\psi^{-}_{j}$ $(j \in \frac{1}{2} + {\Bbb Z}$)
and the following defining relations (i.e. $\psi^{\pm}_{i}$ are free
charged fermions):

$$\psi^{+}_{i} \psi^{-}_{j} + \psi^{-}_{j} \psi^{+}_{i} =
\delta_{i,-j},\ \psi^{\pm}_{i} \psi^{\pm}_{j} + \psi^{\pm}_{j}
\psi^{\pm}_{i} = 0. \tag{0.2.1}$$

The algebra $C\ell$ has a unique irreducible representation in a
vector space $F$ (resp. $F^{*}$) which is a left (resp. right) module
admitting a non-zero vector $| 0\rangle$ (resp. $\langle 0|$)
satisfying

$$\psi^{\pm}_{j}|0 \rangle = 0\ (\text{resp.}\ \langle 0| \psi^{\pm}_{-j}
= 0) \ \text{for}\ j > 0. \tag{0.2.2}$$
These representations are dual to each other with respect to the
pairing

$$\langle \langle 0|a,\ b|0\rangle \rangle = \langle 0|ab|0\rangle$$
normalized by the condition $\langle 0|1|0\rangle = 1$.

The Lie algebra $g\ell_{\infty}$ embeds in $C\ell$ by letting

$$r(E_{ij}) = \psi^{+}_{-i}\psi^{-}_{j}. \tag{0.2.3}$$
Exponentiating gives a representation $R$ of the group $GL_{\infty}$ on
$F$ and $F^{*}$.  Let for $n \in {\Bbb Z}$:

$$\alpha_{n} = \sum_{j \in \frac{1}{2} + {\Bbb Z}}
\psi^{+}_{-j}\psi^{-}_{j+n}   \
\text{for}\ n \neq 0,\ \alpha_{0} = \sum_{j > 0}
\psi^{+}_{-j}\psi^{-}_{j}-\sum_{j < 0} \psi^{-}_{j} \psi^{+}_{-j}.
\tag{0.2.4}$$
and consider the following operator on $F$:

$$H(x) = \sum^{\infty}_{n=1} x_{n}\alpha_{n}. \tag{0.2.5}$$
For a positive integer $m$ let
$$\langle \pm m| \kern-1.1pt = \kern-1.1pt \langle
0|\psi^{\pm}_{\frac{1}{2}} \ldots \psi^{\pm}_{m-\frac{1}{2}} \in
F^{*}\ \text{and}\ |m\rangle \kern-1.1pt = \kern-1.1pt
\psi^{\pm}_{-m+\frac{1}{2}} \ldots
\psi^{\pm}_{-\frac{1}{2}} |0\rangle \in F.$$
Then the Fock space is realized on the vector space of polynomials $B
= {\Bbb C}[x_{1},x_{2},\ldots ;Q,Q^{-1}]$ via the isomorphism $\sigma
:\ F @>\sim>> B$ defined by

$$\sigma (a|0\rangle ) =  \sum_{m \in {\Bbb Z}}
\langle m|e^{H(x)} a|0\rangle Q^{m} . \tag{0.2.6}$$
This remarkable isomorphism is called the boson-fermion
correspondence and goes back to the work of Skyrme [Sk] and many
other physicists; this beautiful form of it is an important part of
the work of Date, Jimbo, Kashiwara and Miwa [DJKM2,3], [JM].

Using that

$$[\alpha_{m},\alpha_{n}] = m\delta_{m,-n}, \tag{0.2.7}$$
(i.e that the $\alpha_{n}$ are free bosons), it is not difficult to
show that the isomorphism $\sigma$ is characterized by the following
two properties [KP2]:

$$\sigma (|m\rangle ) = Q^{m},\ \sigma \alpha_{n} \sigma^{-1} =
\frac{\partial}{\partial x_{n}}\ \text{and}\ \sigma
\alpha_{-n}\sigma^{-1} = nx_{n}\ \text{if}\ n > 0. \tag{0.2.8}$$
Using (0.2.8), it is easy to recover the following well-known
properties of the boson-fermion correspondence [DJKM2,3], [KP2].
Introduce the fermionic fields

$$\psi^{\pm} (z) = \sum_{j \in \frac{1}{2} + {\Bbb Z}}
\psi^{\pm}_{j}z^{-j-1/2}. $$
Then one has:

$$\sigma \psi^{\pm} (z) \sigma^{-1} = Q^{\pm 1}z^{\pm \alpha_{0}}
\Gamma^{\pm}(z), \tag{0.2.9}$$

$$\sigma (\sum_{i,j \in \frac{1}{2} + {\Bbb Z}} r(E_{ij})
z^{i-\frac{1}{2}}_{1} z^{-j-\frac{1}{2}}_{2})\sigma^{-1} =
\frac{1}{z_{1}-z_{2}} \Gamma (z_{1},z_{2}).
\tag{0.2.10}$$
Hence $\Gamma (z_{1},z_{2})$ lies in a ``completion'' of the Lie
algebra $g\ell_{\infty}$ acting on $B$ via the boson-fermion
correspondence.  Therefore, the group $GL_{\infty}$ and its
``completion'' act on $B$ and Date, Jimbo, Kashiwara and Miwa show
that all elements of the orbit ${\Cal O} = GL_{\infty} \cdot 1$ and
its completions satisfy the bilinear identity (0.1.8).  Since $\Gamma
(z_{1},z_{2})^{2} = 0$ and $\Gamma (z_{1},z_{2})$ lies in a
completion of $g\ell_{\infty}$, we see that $\exp a \Gamma
(z_{1},z_{2}) = 1 + a\Gamma (z_{1},z_{2})$ leaves a completion of the orbit
${\Cal O}$ invariant,
which explains why (0.1.15) are solutions of the KP hierarchy.

Since the orbit $GL_{\infty}|0\rangle$ (which is the image of ${\Cal
O}$ in the fermionic picture) can be naturally identified with the
cone over  a
Grassmannian, we arrive at the remarkable discovery of Sato that
solutions of the KP hierarchy are parameterized by an
infinite-dimensional Grassmannian [S].

\vskip 10pt
{\bf 0.3.}  It was subsequently pointed out in [KP2] and [KR]
that the bilinear equation (0.1.8) (in the bosonic picture)
corresponds to the following remarkably simple equation on the
$\tau$-function in the fermionic picture:

$$\sum_{k \in \frac{1}{2} + {\Bbb Z}} \psi^{+}_{k} \tau \otimes
\psi^{-}_{-k} \tau = 0. \tag{0.3.1}$$
This is the fermionic formulation of the KP hierarchy.  Since (0.3.1)
is equivalent to

$$\text{Res}_{z=0} \psi^{+}(z)\tau \otimes \psi^{-}(z)\tau = 0,
\tag{0.3.2}$$
it is clear from (0.1.10 and 11) that equations (0.1.8) and (0.3.2)
are equivalent.  Since $\tau = |0\rangle$ obviously satisfies (0.3.1)
and $R \otimes R(GL_{\infty})$ commutes with the operator $\sum_{k}
\psi^{+}_{k} \otimes \psi^{-}_{-k}$, we see why any element of
$R(GL_{\infty})|0\rangle$ satisfies (0.3.1).  Thus, the most natural
approach to the KP hierarchy is to start with the fermonic
formulation (0.3.1), go over to the bilinear identity (0.1.8) and
then to all other formulations (see [KP2], [KR], [K]).  This approach
was generalized in [KW].

\vskip 10pt
{\bf 0.4.}  Our basic idea is to start once again with the fermonic
formulation of KP, but then use the $n$-component boson-fermion
correspondence, also considered by Date, Jimbo, Kashiwara and Miwa
[DJKM1,2], [JM].  This leads to a bilinear equation on a matrix wave function,
which in turn leads to a deformation equation for a matrix formal
pseudo-differential operator, to matrix Sato equations and to matrix
Zakharov-Shabat type equations.

The corresponding linear problem has been already formulated in
Sato's paper [S] and Date, Jimbo, Kashiwara and Miwa [DJKM1]
have written the corresponding bilinear equation for the wave
function, but the connection between these formulations remained
somewhat obscure.

It is the aim of the present paper to give all formulations of the
$n$-component KP hierarchy and clarify connections between them.
The generalization to the $n$-component KP is important because it
contains many of the most popular systems of soliton equations, like
the Davey-Stewartson system (for $n = 2$), the $2$-dimensional Toda
lattice (for $n = 2$), the $n$-wave system (for $n \geq 3$).  It
also allows us to construct natural generalizations of the
Davey-Stewartson and Toda lattice systems.  Of course, the inclusion
of all these systems in the $n$-component KP hierarchy allows us to
construct their solutions by making use of vertex operators.

Hirota's direct method [H] requires some guesswork to introduce a new
function (the $\tau$-function) for which the equations in question
take a bilinear form.  The inclusion of the equations in the
$n$-component KP hierarchy provides a systematic way of construction
of the $\tau$-functions, the corresponding bilinear equations and a
large family of their solutions.

The difficulty of the $\tau$-function approach lies in the fact that
the hierarchy contains too many Hirota bilinear equations.  To deal
with this difficulty we introduce the notion of an energy of a Hirota
bilinear equation.  We observe that the most interesting equations
are those of lowest energy.  For example, in the $n = 1$ case the
lowest energy $(= 4)$ non-trivial equation is the classical KP
equation in the Hirota bilinear form, in the $n = 2$ case
the lowest energy $(= 2)$ equations form the $2$-dimensional Toda
chain and the energy $2$ and $3$ equations form the Davey-Stewartson
system in the bilinear
form, and in the  $n \geq 3$ case the lowest energy $(= 2)$ bilinear
equations form the $n$-wave system in the bilinear form.

There is a new phenomenon in the $n$-component case, which does not
occur in the $1$-component case: the $\tau$-function and the
wave function is a collection of functions $\{ \tau_{\alpha}\}$ and
$\{ W_{\alpha}\}$ parameterized by the elements of the root lattice
$M$ of type $A_{n-1}$.  The set $\text{supp}\ \tau = \{ \alpha \in
M|\tau_{\alpha} \neq 0\}$ is called the support of the
$\tau$-function $\tau$.  We show that $\text{supp} \ \tau$ is a
convex polyhedron whose edges are parallel to roots; in particular,
$\text{supp}\ \tau$ is connected, which allows us to relate the
behaviour of the $n$-component KP hierarchy at different points of
the lattice $M$.  It is interesting to note that the ``matching
conditions'' which relate the functions $W_{\alpha}$ and
$W_{\beta}$, $\alpha , \beta \in M$, involve elements from the
subgroup of translations of the Weyl group [K, Chapter 6] of the loop
group $GL({\Bbb C}[z,z^{-1}])$ and are intimately related to the
Bruhat decomposition of this loop group (see [PK]).  We are planning
to study this in a future publication.

The behaviour of solutions obtained via vertex operators in the
$n$-component case is much more complicated than for the ordinary KP
hiearchy.  In particular, they are not necessarily multisoliton
solutions (i.e. a collection of waves that preserve their form after
interaction).  For that reason we call them the multisolitary
solutions.  Some of the multisolitary solutions turn out to be the so
called dromion solutions, that have become very popular recently
[BLMP], [FS], [HH], [HMM].  This solutions decay exponentially in all
directions (and they are not soliton solutions; in particular, they
exist only for $n > 1$).  It is a very interesting problem for which
values of parameters the multisolitary solutions are soliton or
dromion solutions.

Note also that the Krichever method for construction of the
quasiperiodic solutions of the KP hierarchy as developed in [SW] and
[Sh] applies to the $n$-component KP.

As shown in [S], [DJKM2], the $m$-th reduction of the KP hiearchy,
i.e. the requirement that $L^{m}$ is a differential operator, leads
to the classical formulation of the celebrated KdV hierarchy for $m =
2$, Boussinesq for $m = 3$ and all the Gelfand-Dickey hierarchies
for $m > 3$.  The totality of $\tau$-functions for the $m$-th
reduced KP hierarchy turns out to be the orbit of the vacuum under
the loop group of $SL_{m}$.

We define in a similar way the $m$-th reduction of the
$n$-component KP and show that the totality of $\tau$-functions is
the orbit of the vacuum vector under the loop group of $SL_{mn}$.
Even the case $m = 1$ turns out to be extremely interesting (it is
trivial if $n = 1$), as it gives the $1+1$ $n$-wave system for $n
\geq 3$ and the decoupled non-linear Schr\"{o}dinger (or AKNS)
system for $n = 2$.   We note that the $1$-reduced $n$-component
KP, which we call
the $n$-component NLS hierarchy, admits a natural generalization to
the case of an arbitrary simple Lie group $G$ (the $n$-component NLS
corresponding to $GL_{n}$).  These hierarchies which might be called
the GNLS hierarchies, contain the systems studied by many authors
[Di], [W1 and 2], [KW],$\ldots$ .

{\bf 0.5.}  The paper is set out as follows.  In \S 1 we explain the
construction of the semi-infinite wedge representation $F$ of the
group $GL_{\infty}$ and write down the equation of the
$GL_{\infty}$-orbit ${\Cal O}$ of the vacuum $|0\rangle$
(Proposition 1.3).  This equation is called the KP hierarchy in the
fermionic picture.  As usual, the Pl\"{u}cker map
makes ${\Cal O}$ a ${\Bbb C}^{\times}$-bundle over an
infinite-dimensional Grassmannian.  We describe the ``support'' of
$\tau \in {\Cal O}$ (Proposition 1.4).

In \S 2 we introduce the $n$-component bosonisation and write down
the fermionic fields in terms of bosonic ones via vertex operators
(Theorem 2.1).  This allows us to transport the KP hierarchy from the
fermionic picture to the bosonic one (2.3.3) and write down the
$n$-component KP hierarchy as a system of Hirota bilinear equations
(2.3.7).  We describe the support of a $\tau$-function in the
bosonic picture (Proposition 2.4).  At the end of the section we list
all Hirota bilinear equations of lowest energy (2.4.3--9).

We start \S 3 with an exposition of the theory of matrix formal
pseudo-differential operators, and prove the crucial Lemma 3.2.
This allows us to reformulate the $n$-component KP hierarchy (2.3.3)
in terms of formal pseudo-differential operators (see (3.3.4 and
12)).  Using the crucial lemma we show that the bilinear equation
(2.3.3) is equivalent to the Sato equation (3.4.2) and matching
conditions (3.3.16) on the wave operators $P^{+}(\alpha )$.  We show
that Sato equation is the compatibility condition of Sato's linear
problem (3.5.5) on the wave function (Proposition 3.5), and that
compatibility of Sato equation implies the equivalent Lax and
Zakharov-Shabat equations (Lemma 3.6).  We prove that compatibility
conditions completely determine the wave operators $P^{+}(\alpha )$
once one of them is given (Proposition 3.3).  At the end of the
section we write down explicitly the first Sato and Lax equations and
relations between them.

In \S 4 we show that many well-known $2+1$ soliton equations are the
simplest equations of the $n$-component KP hierarchy, and deduce
from \S 3 expressions for their $\tau$-functions and the
corresponding Hirota bilinear equations.

Using vertex operators we write down in \S 5 the $N$-solitary
solutions (5.1.11) of the $n$-component KP and hence of all its
relatives.  We discuss briefly the relation of this general solution
to the known solutions to the relatives.

In \S 6 we discuss the $m$-reductions of the $n$-component KP
hierarchy.  They reduce the $2+1$ soliton equations to $1+1$ soliton
equations.  We show that at the group theoretic level it corresponds
to a reduction from $GL_{\infty}$ (or rather a completion of it) to
the subgroup $SL_{mn}({\Bbb C}[t,t^{-1}])$ (Proposition 6.1).  We
discuss in more detail the $1$-reduced $n$-component KP, which is a
generalization of the NLS system and which admits further
generalization to any simple Lie group.

We would like to thank A.S. Fokas for asking one of us to write a
paper for this volume.  We are grateful to E. Medina for calling our
attention to the paper [HMM].  The second named author would like to
thank MIT for the kind hospitality.

\subheading{\S 1.  The semi-infinite wedge
representation of the group $GL_{\infty}$ and the KP hierarchy in
the fermionic picture}

\vskip 10pt
{\bf 1.1.}  Consider the infinite complex matrix group

$$GL_{\infty} = \{ A = (a_{ij})_{i,j \in {\Bbb Z}+\frac{1}{2}}|A\
\text{is invertible and all but a finite number of}\ a_{ij} -
\delta_{ij}\ \text{are}\ 0\}$$
and its Lie algebra

$$gl_{\infty} = \{ a = (a_{ij})_{i,j \in {\Bbb Z}+\frac{1}{2}}|\
\text{all but a finite number of}\ a_{ij}\ \text{are}\ 0\}$$
with bracket $[a,b] = ab-ba$.  The Lie algebra $gl_{\infty}$ has a
basis consisting of matrices $E_{ij},\ i,j \in {\Bbb Z} + \frac{1}{2}$, where
$E_{ij}$ is the matrix with a $1$ on the $(i,j)$-th entry and zeros
elsewhere.

Let ${\Bbb C}^{\infty} = \bigoplus_{j \in {\Bbb Z}+\frac{1}{2}} {\Bbb
C} v_{j}$ be an infinite dimensional complex vector space with fixed
basis $\{ v_{j}\}_{j \in {\Bbb Z}+\frac{1}{2}}$.  Both the group
$GL_{\infty}$ and its Lie algebra $gl_{\infty}$ act linearly on
${\Bbb C}^{\infty}$ via the usual formula:

$$E_{ij} (v_{k}) = \delta_{jk} v_{i}.$$

\vskip 10pt
  The well-known semi-infinite wedge representation is
constructed as follows [KP2].  The semi-infinite wedge space $F =
\Lambda^{\frac{1}{2}\infty} {\Bbb C}^{\infty}$ is the vector space
with a basis consisting of all semi-infinite monomials of the form
$v_{i_{1}} \wedge v_{i_{2}} \wedge v_{i_{3}} \ldots$, where $i_{1} >
i_{2} > i_{3} > \ldots$ and $i_{\ell +1} = i_{\ell} -1$ for $\ell >>
0$.  We can now define representations $R$ of $GL_{\infty}$ and $r$
of $gl_{\infty}$ on $F$ by

$$
R(A) (v_{i_{1}} \wedge v_{i_{2}} \wedge v_{i_{3}} \wedge \cdots) = A
v_{i_{1}} \wedge Av_{i_{2}} \wedge Av_{i_{3}} \wedge \cdots , \tag{1.1.1}
$$

$$r(a) (v_{i_{1}} \wedge v_{i_{2}} \wedge v_{i_{3}} \wedge \cdots ) =
\sum_{k} v_{i_{1}} \wedge v_{i_{2}} \wedge \cdots \wedge v_{i_{k-1}}
\wedge av_{i_{k}} \wedge v_{i_{k+1}} \wedge \cdots . \tag{1.1.2}
$$
These equations are related by the usual formula:

$$\exp (r(a)) = R(\exp a)\ \text{for}\ a \in gl_{\infty}.$$

\vskip 10pt
{\bf 1.2.}  The representation $r$ of $gl_{\infty}$ can be described
in  terms of a Clifford algebra.  Define the wedging and contracting
operators $\psi^{+}_{j}$ and $\psi^{-}_{j}\ \ (j \in {\Bbb Z} +
\frac{1}{2})$ on $F$ by

$$\align
&\psi^{+}_{j} (v_{i_{1}} \wedge v_{i_{2}} \wedge \cdots ) = \cases 0
& \text{if}\ j = i_{s} \text{for some}\ s \\
(-1)^{s} v_{i_{1}} \wedge v_{i_{2}} \cdots \wedge v_{i_{s}} \wedge
v_{-j} \wedge v_{i_{s+1}} \wedge \cdots &\text{if}\ i_{s} > -j >
i_{s+1}\endcases \\
&\psi^{-}_{j} (v_{i_{1}} \wedge v_{i_{2}} \wedge \cdots ) = \cases 0
&\text{if}\ j \neq i_{s}\ \text{for all}\ s \\
(-1)^{s+1} v_{i_{1}} \wedge v_{i_{2}} \wedge \cdots \wedge
v_{i_{s-1}} \wedge v_{i_{s+1}} \wedge \cdots &\text{if}\ j = i_{s}.
\endcases
\endalign
$$
These operators satisfy the following relations
$(i,j \in {\Bbb Z}+\frac{1}{2}, \lambda ,\mu = +,-)$:

$$\psi^{\lambda}_{i} \psi^{\mu}_{j} + \psi^{\mu}_{j}
\psi^{\lambda}_{i} = \delta_{\lambda ,-\mu} \delta_{i,-j}, \tag{1.2.1}$$
hence they generate a Clifford algebra, which we denote by ${\Cal C}\ell$.

Introduce the following elements of $F$ $(m \in {\Bbb Z})$:

$$|m\rangle = v_{m-\frac{1}{2} } \wedge v_{m-\frac{3}{2} } \wedge
v_{m-\frac{5}{2} } \wedge \cdots .$$
It is clear that $F$ is an irreducible ${\Cal C}\ell$-module such that

$$\psi^{\pm}_{j} |0\rangle = 0 \ \text{for}\ j > 0 . \tag{1.2.2}$$

It is straightforward that the representation $r$ is given by the
following formula:

$$r(E_{ij}) = \psi^{+}_{-i} \psi^{-}_{j}. \tag{1.2.3}$$

Define the {\it charge decomposition}

$$F = \bigoplus_{m \in {\Bbb Z}} F^{(m)}$$
by letting

$$\text{charge}(v_{i_{1}} \wedge v_{i_{2}} \wedge \ldots ) = m\
\text{if} \ i_{k} + k = {\ssize \frac{1}{2}} + m \ \text{for}\ k >> 0.
\tag{1.2.5}$$
Note that
$$\text{charge}(|m\rangle ) = m\ \text{and charge} (\psi^{\pm}_{j}) =
\pm 1. \tag{1.2.6}$$

It is clear that the charge decomposition is invariant with respect
to $r(g\ell_{\infty})$ (and hence with respect to $R(GL_{\infty})$).
Moreover, it is easy to see that each $F^{(m)}$ is irreducible with
respect to $g\ell_{\infty}$ (and $GL_{\infty}$).  Note that
$|m\rangle$ is its highest weight vector, i.e.

$$\align
&r(E_{ij})|m\rangle = 0 \ \text{for}\ i < j, \\
&r(E_{ii})|m\rangle = 0\  (\text{resp.}\ = |m\rangle ) \ \text{if}\ i > m\
(\text{resp. if}\ i < m).
\endalign
$$

\vskip 10pt
{\bf 1.3.}  The main object of our study is the
$GL_{\infty}\text{-orbit}$
$${\Cal O}
= R(GL_{\infty})|0\rangle \subset F^{(0)}$$
of the vacuum vector $|0\rangle$.

\proclaim{Proposition 1.3 ([KP2])}  A non-zero element $\tau$ of $F^{(0)}$
lies in ${\Cal O}$ if and only if the following equation holds in $F \otimes
F$:

$$\sum_{k \in {\Bbb Z}+\frac{1}{2}} \psi^{+}_{k} \tau \otimes \psi^{-}_{-k}
\tau = 0. \tag{1.3.1}$$
\endproclaim

\demo{Proof} It is clear that $\sum_{k} \psi^{+}_{k} |0\rangle \otimes
\psi^{-}_{-k} |0\rangle = 0$ and it is easy to see that the operator $\sum_{k}
\psi^{+}_{k} \otimes \psi^{-}_{k} \in \text{End}(F\otimes F)$ commutes with
$R(g) \otimes R(g)$ for any $g \in GL_{\infty}$.  It follows that
$R(g)|0\rangle$ satisfies (1.3.1).  For the proof of the converse statement
(which is not important for our purposes) see [KP2] or [KR].\ \ \ $\square$
\enddemo

Equation (1.3.1) is called the {\it KP hierarchy in the fermionic
picture}.

Note that any non-zero element $\tau$ from the orbit ${\Cal O}$ is
of the form:

$$\tau = u_{-\frac{1}{2}} \wedge u_{-\frac{3}{2}} \wedge
u_{-\frac{5}{2}} \wedge  \ldots
,\text{where}\  u_{j} \in
{\Bbb C}^{\infty}\ \text{and}\ u_{-k} = v_{-k}\ \text{for}\ k >\!\!>
0. \tag{1.3.2}$$
This allows us to construct a canonical map $\varphi :\ {\Cal O} @>>>
\ \text{Gr}$ by $\varphi (\tau ) = \sum_{i} {\Bbb C} u_{-i}
\subset {\Bbb C}^{\infty}$, where $\text{Gr}$  consists of the
subspaces of ${\Bbb C}^{\infty}$ containing $\sum^{\infty}_{j=k} {\Bbb
C}v_{-j-1/2}$ for $k >\!\!> 0$ as a subspace of codimension $k$.  It is
clear that the map $\varphi$ is surjective with fibers ${\Bbb
C}^{\times}$.

\vskip 10pt
{\bf 1.4.}  Consider the free ${\Bbb Z}$-module ${\tilde L}$ with
the basis  $\{\delta_{j}\}_{j \in \frac{1}{2}+{\Bbb Z}}$, let
${\tilde \Delta}$ (resp. ${\tilde \Delta}_{0}$) $= \{ \delta_{i} -
\delta_{j}|i,j \in \frac{1}{2} +
{\Bbb Z}$ (resp. $i,-j \in \frac{1}{2} + {\Bbb Z}_{+}$), $i \neq j\}$, and
let ${\tilde M} \subset {\tilde L}$ (resp. ${\tilde M}_{0} \subset
{\tilde L}$) be the ${\Bbb Z}$-span of ${\tilde \Delta}$ (resp. ${\tilde
\Delta}_{0}$). We define the weight of a semi-infinite monomial by

$$\text{weight} (\psi^{+}_{i_{1}} \ldots
\psi^{+}_{i_{s}}\psi^{-}_{j_{1}} \ldots \psi^{-}_{j_{t}}|0\rangle ) =
\delta_{-i_{1}} + \ldots + \delta_{-i_{s}} - \delta_{j_{1}} - \ldots
- \delta_{j_{t}}. \tag{1.4.1}$$
Note that weights of semi-infinite monomials from $F^{(0)}$ lie in
${\tilde M}_{0}$.  Given
$\tau \in F$ we denote by $f \text{supp}\, \tau $, and call it the
{\it fermionic support of} $\tau$, the set of weights of semi-infinite
monomials that occur in $\tau$ with a non-zero coefficient.

\proclaim{Proposition 1.4}  If $\tau \in {\Cal O}$, then
$f\text{supp}\ \tau$ is a cube consisting of all sums of elements of
a finite subset of ${\tilde \Delta}_{0}$.
\endproclaim

\demo{Proof} According to the general result [PK, Lemma 4], the edges
of the convex hull of $f\text{supp}\ \tau$ must be parallel to the
elements of ${\tilde \Delta}_{0}$.  But if the difference of weights of two
semi-infinite monomials
is a multiple of $\delta_{i}-\delta_{j}$, then it is clearly equal to
$\pm(\delta_{i}-\delta_{j})$. Hence edges of the convex hull of
$f\text{supp}\ \tau$ are translations of elements of
${\tilde \Delta}_{0}$, and the proposition follows. \ \ \ $\square$
\enddemo

\subheading{\S 2. The $n$-component bosonization and the KP hierarchy in the
bosonic picture}

\vskip 10pt
{\bf 2.1.}  Using a bosonization one can rewrite (1.3.1) as a system
of partial
differential equations.  There are however many different bosonizations.  In
this paper we focus on the $n$-component bosonizations, where $n = 1,2,\ldots
$.

For that purpose we relabel the basis vectors $v_{i}$ and with them the
corresponding fermionic operators (the wedging and contracting operators).
This relabeling can be done in many different ways, see e.g. [TV], the
simplest one is the following.

Fix $n \in {\Bbb N}$ and define for $j \in {\Bbb Z},\ 1 \leq j \leq n,\ k \in
{\Bbb Z} + \frac{1}{2}$:

$$v^{(j)}_{k} = v_{nk - \frac{1}{2}(n-2j+1)},$$
and correspondingly:

$$\psi^{\pm (j)}_{k} = \psi^{\pm}_{nk \pm \frac{1}{2}
(n-2j+1)}.$$
Notice that with this relabeling we have:

$$\psi^{\pm (j)}_{k}|0\rangle = 0\ \text{for}\ k > 0.$$

The charge decomposition (1.2.5) can be further decomposed into a sum
of {\it partial charges} which are denoted by $\text{charge}_{j}$, $j =
1,\ldots ,n$,  defined for a semi-infinite
monomial $v \equiv v_{i_{1}} \wedge v_{i_{2}} \wedge \ldots$ of
weight $\sum_{i} a_{i} \delta_{i}$ by

$$\text{charge}_{j} (v) = \sum_{k \in {\Bbb Z}} a_{kn + j-1/2}.
\tag{2.1.1}$$
Another important decomposition is the {\it energy decomposition}
defined by

$$\text{energy}\ |0\rangle = 0,\ \text{energy}\ \psi^{\pm (j)}_{k} =
-k. \tag{2.1.2}$$
Note that energy is a non-negative number which can be calculated by

$$\text{energy} (v) = \sum_{k \in \frac{1}{2} + {\Bbb Z}} a_{k}
( [ k/n ]  + {\tsize {\frac{1}{2} }}).
\tag{2.1.3}$$

Introduce the  fermionic fields $(z \in {\Bbb C}^{\times})$:

$$\psi^{\pm (j)}(z) \overset{\text{def}}\to{=} \sum_{k \in {\Bbb
Z}+\frac{1}{2}} \psi^{\pm
(j)}_{k} z^{-k-\frac{1}{2}}.\tag{2.1.4}$$
Next we introduce bosonic fields $(1 \leq i,j \leq n)$:

$$\alpha^{(ij)}(z) \equiv \sum_{k \in {\Bbb Z}} \alpha^{(ij)}_{k} z^{-k-1}
\overset{def}\to{=} :\psi^{+(i)}(z) \psi^{-(j)}(z):, \tag{2.1.5}$$
where $:\ :$ stands for the {\it normal ordered product} defined in
the usual way $(\lambda ,\mu = +$ or $-$):

$$:\psi^{\lambda (i)}_{k} \psi^{\mu (j)}_{\ell}: = \cases \psi^{
\lambda (i)}_{k}
\psi^{\mu (j)}_{\ell}\ &\text{if}\ \ell > 0 \\
-\psi^{\mu (j)}_{\ell} \psi^{\lambda (i)}_{k} &\text{if}\ \ell <
0.\endcases \tag{2.1.6}$$
One checks (using e.g. the Wick formula) that the operators
$\alpha^{(ij)}_{k}$ satisfy the commutation relations of the affine
algebra $gl_{n}({\Bbb C})^{\wedge}$ with central charge $1$, i.e.:

$$[\alpha^{(ij)}_{p},\alpha^{(k\ell)}_{q}] =
\delta_{jk}\alpha^{(i\ell )}_{p+q}
- \delta_{i\ell} \alpha^{(jk)}_{p+q} + p\delta_{i\ell}
\delta_{jk}\delta_{p,-q},\tag{2.1.7}$$
and that
$$\alpha^{(ij)}_{k}|m\rangle = 0 \ \text{if}\ k > 0 \ \text{or}\ k = 0\
\text{and}\ i < j.\tag{2.1.8}$$
The operators $\alpha^{(i)}_{k} \equiv \alpha^{(ii)}_{k}$
satsify the canonical commutation relation of the associative
oscillator algebra,  which we
denote by ${\frak a}$:

$$[\alpha^{(i)}_{k},\alpha^{(j)}_{\ell}] =
k\delta_{ij}\delta_{k,-\ell},\tag{2.1.9}$$
and one has

$$\alpha^{(i)}_{k}|m\rangle = 0 \ \text{for}\ k > 0.\tag{2.1.10}$$

It is easy to see that restricted to $g\ell_{n}({\Bbb C})^{\wedge}$,
$F^{(0)}$ is its basic highest weight representation (see [K, Chapter
12]).  The $g\ell_{n}({\Bbb C})^{\wedge}$-weight of a
semi-infinite monomial $v$ is as follows:

$$\Lambda_{0} + \sum^{n}_{j=1} \ \text{charge}_{j} (v) \delta_{j} - \
\text{energy} (v){\tilde \delta}. \tag{2.1.11}$$
Here $\Lambda_{0}$ is the highest weight of the basic representation,
$\{ \delta_{j}\}$ is the standard basis of the weight lattice of
$g\ell_{n}({\Bbb C})$ and ${\tilde \delta}$ is the primitive
imaginary root ([K, Chapter 7]).

In order to express the fermionic fields $\psi^{\pm (i)}(z)$ in terms of
the bosonic fields $\alpha^{(ii)}(z)$, we need some additional operators
$Q_{i},\ i = 1,\ldots ,n$, on $F$.  These operators are uniquely defined by
the following conditions:

$$Q_{i}|0\rangle = \psi^{+(i)}_{-\frac{1}{2}} |0\rangle ,\ Q_{i}\psi^{\pm
(j)}_{k} = (-1)^{\delta_{ij}+1} \psi^{\pm
(j)}_{k\mp \delta_{ij}}Q_{i}.\tag{2.1.12}$$
They satisfy the following commutation relations:

$$Q_{i}Q_{j} = -Q_{j}Q_{i}\ \text{if}\ i \neq j,\ [\alpha^{(i)}_{k},Q_{j}] =
\delta_{ij} \delta_{k0}Q_{j}.\tag{2.1.13}$$

\proclaim{Theorem 2.1}  ([DJKM1], [JM])
$$\psi^{\pm (i)}(z) = Q^{\pm 1}_{i}z^{\pm \alpha^{(i)}_{0}} \exp
(\mp \sum_{k < 0} \frac{1}{k} \alpha^{(i)}_{k}z^{-k})\exp(\mp
\sum_{k > 0} \frac{1}{k} \alpha^{(i)}_{k} z^{-k}). \tag{2.1.14}$$
\endproclaim

\demo{Proof} see [TV].
\enddemo

The operators on the right-hand side of (2.1.14) are called vertex
operators.  They made their first appearance in string theory (cf.
[FK]).

We shall use below the following notation

$$|k_{1},\ldots ,k_{n}\rangle = Q^{k_{1}}_{1} \ldots
Q^{k_{n}}_{n}|0\rangle . \tag{2.1.15}$$
\vskip 10pt

REMARK 2.1.  One easily checks the following relations:

$$[\alpha^{(i)}_{k},\psi^{\pm (j)}_{m}] = \pm \delta_{ij} \psi^{\pm
(j)}_{k+m} .$$
They imply formula (2.1.14) for $\psi^{\pm (i)}(z)$ except for the
first two factors, which require some additional analysis.

\vskip 10pt
{\bf 2.2.}  We can describe now the $n$-component boson-fermion
correspondence.  Let ${\Bbb C}[x]$ be the space of polynomials in
indeterminates $x = \{ x^{(i)}_{k}\},\ k = 1,2,\ldots ,\ i =
1,2,\ldots ,n$.  Let $L$ be a lattice with a basis $\delta_{1},\ldots
,\delta_{n}$ over ${\Bbb Z}$ and the symmetric bilinear form
$(\delta_{i}|\delta_{j}) = \delta_{ij}$, where $\delta_{ij}$ is the
Kronecker symbol.  Let

$$\varepsilon_{ij} = \cases -1 &\text{if $i > j$} \\
1 &\text{if $i \leq j$.} \endcases \tag{2.2.1}$$

Define a bimultiplicative function $\varepsilon :\ L \times L @>>> \{
\pm 1 \}$ by letting

$$\varepsilon (\delta_{i}, \delta_{j}) = \varepsilon_{ij}.
\tag{2.2.2}$$
Let $\delta = \delta_{1} + \ldots + \delta_{n},\ M = \{ \gamma \in
L|\ (\delta | \gamma ) = 0\}$, $\Delta = \{ \alpha_{ij} :=
\delta_{i}-\delta_{j}| i,j = 1,\ldots ,n,\ i \neq j \}$.  Of course
$M$ is the root lattice of $s\ell_{n}({\Bbb C})$, the set $\Delta$
being the root system.

Consider the vector space ${\Bbb C}[L]$ with basis $e^{\gamma}$,\
$\gamma \in L$, and the following twisted group algebra product:

$$e^{\alpha}e^{\beta} = \varepsilon (\alpha ,\beta)e^{\alpha +
\beta}. \tag{2.2.3}$$
Let $B = {\Bbb C}[x] \otimes_{\Bbb C} {\Bbb C}[L]$ be the tensor
product of algebras.  Then the $n$-component boson-fermion
correspondence is the vector space isomorphism

$$\sigma :F @>\sim >> B, \tag{2.2.4}$$
given by

$$\sigma (\alpha^{(i_{1})}_{-m_{1}} \ldots
\alpha^{(i_{s})}_{-m_{s}}|k_{1},\ldots ,k_{n}\rangle ) = m_{1} \ldots
m_{s}x^{(i_{1})}_{m_{1}} \ldots x^{(i_{s})}_{m_{s}} \otimes
e^{k_{1}\delta_{1} + \ldots + k_{n}\delta_{n}} . \tag{2.2.5}$$

The transported charge and energy then will be as follows:

$$\align
&\text{charge}(p(x)\otimes e^{\gamma}) = (\delta |\gamma),
\ \text{charge}_{j}(p(x) \otimes e^{\gamma}) = (\delta_{j}|\gamma),
\tag{2.2.6}\\
&\text{energy}(x^{(i_{1})}_{m_{1}} \ldots x^{(i_{s})}_{m_{s}} \otimes
e^{\gamma}) = m_{1} + \ldots + m_{s} + {\ssize \frac{1}{2}} (\gamma |\gamma).
\tag{2.2.7}
\endalign$$
We denote the transported charge decomposition by

$$B = \bigoplus_{m \in {\Bbb Z}} B^{(m)}.$$

The transported action of the operators $\alpha^{(i)}_{m}$ and $Q_{j}$ looks
as follows:

$$\cases
\sigma \alpha^{(j)}_{-m}\sigma^{-1}(p(x) \otimes e^{\gamma}) =
mx^{(j)}_{m}p(x)\otimes e^{\gamma},\ \text{if}\ m > 0, &\  \\
\sigma \alpha^{(j)}_{m} \sigma^{-1}(p(x) \otimes e^{\gamma}) = \frac{\partial
p(x)}{\partial x_{m}} \otimes e^{\gamma},\ \text{if}\ m > 0, &\  \\
\sigma \alpha^{(j)}_{0} \sigma^{-1} (p(x) \otimes e^{\gamma}) =
(\delta_{j}|\gamma ) p(x) \otimes e^{\gamma} , &\ \\
\sigma Q_{j} \sigma^{-1} (p(x) \otimes e^{\gamma}) = \varepsilon
(\delta_{j},\gamma)  p(x) \otimes e^{\gamma + \delta_{j}}
. & \
\endcases \tag{2.2.8}
$$

\vskip 10pt
{\bf 2.3.}  Using the isomorphism $\sigma$ we can reformulate the KP hierarchy
(1.3.1) in the bosonic picture as a hierarchy of Hirota bilinear equations.

We start by observing that (1.3.1) can be rewritten as follows:

$$\text{Res}_{z=0}\ dz ( \sum^{n}_{j=1} \psi^{+(j)}(z)\tau
\otimes \psi^{-(j)}(z)\tau ) = 0,\ \tau \in F^{(0)}.
\tag{2.3.1}$$
Here and further $\text{Res}_{z=0}\ dz \sum_{j} f_{j}z^{j}$ (where
$f_{j}$ are independent of $z$) stands for $f_{-1}$.
Notice that for $\tau \in F^{(0)},\ \sigma (\tau) = \sum_{\gamma \in M}
\tau_{\gamma}(x)e^{\gamma}$.
 Here and further  we write $\tau_{\gamma}(x)e^{\gamma}$ for
$\tau_{\gamma} \otimes
e^{\gamma}$.  Using Theorem 2.1, equation (2.3.1) turns under $\sigma
\otimes \sigma :\ F \otimes F \overset\,\,\sim\to\longrightarrow
{\Bbb C}[x^{\prime},x^{\prime \prime}]
\otimes ({\Bbb C}[L^{\prime}] \otimes {\Bbb C}[L^{\prime \prime}])$ into the
following equation:
$$
\aligned
&\text{Res}_{z=0}\ dz (
\sum^{n}_{j=1} \sum_{\alpha ,\beta
\in M} \varepsilon (\delta_{j},\alpha - \beta)
z^{(\delta_{j}|\alpha -\beta)} \\
& \times \exp (\sum^{\infty}_{k=1}
(x^{(j)^{\prime}}_{k} - x^{(j)^{\prime \prime}}_{k} )z^{k}) \exp
( -\sum^{\infty}_{k=1} ( \frac{\partial}{\partial
x^{(j)^{\prime}}_{k}} -
\frac{\partial}{\partial x^{(j)^{\prime \prime}}_{k}} ) \frac{z^{-k}}{k}
) \\
& \tau_{\alpha}(x^{\prime})(e^{\alpha + \delta_{j}})^{\prime}
\tau_{\beta}(x^{\prime \prime})(e^{\beta - \delta_{j}})^{\prime
\prime}) = 0  .
\endaligned \tag{2.3.2}$$
Hence for all $\alpha ,\beta \in L$ such that $(\alpha
|\delta ) = -(\beta |\delta ) = 1$ we have:

$$\aligned
&\text{Res}_{z=0} ( dz
 \sum^{n}_{j=1} \varepsilon (\delta_{j}, \alpha-\beta)
z^{(\delta_{j}|\alpha - \beta - 2\delta_{j})}  \\
 &\times \exp
(\sum^{\infty}_{k=1} (x^{(j)^{\prime}}_{k} - x^{(j)^{\prime
\prime}}_{k})z^{k})
\exp (-\sum^{\infty}_{k=1} (\frac{\partial}{\partial x^{(j)^{\prime}}_{k}}
 - \frac {\partial}{\partial x^{(j)^{\prime
\prime}}_{k}})\frac{z^{-k}}{k}) \\
& \tau_{\alpha -
\delta_{j}}(x^{\prime})\tau_{\beta + \delta_{j}}(x^{\prime \prime}))
= 0  . \endaligned \tag{2.3.3}$$
Now making the change of variables
$$x^{(j)}_{k} = {\ssize \frac{1}{2}} (x^{(j)^{\prime}}_{n} + x^{(j)^{\prime
\prime}}_{n}), \quad y^{(j)}_{k} = {\ssize \frac{1}{2}} (x^{(j)^{\prime}}_{k} -
x^{(j)^{\prime \prime}}_{n}),$$
(2.3.3) becomes

$$\aligned
&\text{Res}_{z=0} ( dz \sum^{n}_{j=1}
\varepsilon (\delta_{j},\alpha -\beta)
z^{(\delta_{j}|\alpha - \beta - 2\delta_{j})}  \\
& \times \exp (\sum^{\infty}_{k=1} 2y^{(j)}_{k} z^{k}) \exp
(-\sum^{\infty}_{k=1}
\frac{\partial}{\partial y^{(j)}_{k}} \frac{z^{-k}}{k}) \tau_{\alpha -
\delta_{j}}(x+y) \tau_{\beta + \delta_{j}} (x-y)) = 0 .
\endaligned \tag{2.3.4}
$$

We can rewrite (2.3.4) using the elementary Schur polynomials defined
by (0.1.13):

$$
\sum^{n}_{j=1} \varepsilon (\delta_{j},\alpha - \beta)
 \sum^{\infty}_{k=0} S_{k}(2y^{(j)})S_{k-1+(\delta_{j}|\alpha
-\beta )} (-\frac{\tilde \partial}{\partial y^{(j)}})
 \tau_{\alpha - \delta_{j}} (x+y) \tau_{\beta + \delta_{j}}
(x-y) = 0.
\tag{2.3.5}$$
Here and further we use the notation

$$\frac{\tilde \partial}{\partial y} = (\frac{\partial}{\partial y_{1}} ,
\frac{1}{2} \frac{\partial}{\partial  y_{2}},\ \frac{1}{3}
\frac{\partial}{\partial y_{3}}, \ldots )$$
Using Taylor's formula we can rewrite (2.3.5) once more:

$$\aligned
&\sum^{n}_{j=1} \varepsilon (\delta_{j},\alpha - \beta) \sum^{\infty}_{k=0}
S_{k}(2y^{(j)})S_{k-1+(\delta_{j}|\alpha -
\beta)} (-\frac{\tilde \partial}{\partial u^{(j)}}) \\
& \times e^{\sum^{n}_{j=1}
\sum^{\infty}_{r=1}y^{(j)}_{r}\frac{\partial}{\partial
u^{(j)}_{r}} } \tau_{\alpha - \delta_{j}} (x+u)\tau_{\beta +
\delta_{j}}(x-u)|_{u=0} = 0.
\endaligned \tag{2.3.6}$$
This last equation can be written as the following generating series of Hirota
bilinear equations:

$$\aligned
&\sum^{n}_{j=1} \varepsilon (\delta_{j} , \alpha - \beta) \sum^{\infty}_{k=0}
S_{k}(2y^{(j)})S_{k-1+(\delta_{j}|\alpha -
\beta)} (-\widetilde{D^{(j)}}) \\
&\times e^{\sum^{n}_{j=1} \sum^{\infty}_{r=1} y^{(j)}_{r}D^{(j)}_{r}}
\tau_{\alpha -
\delta_{j}} \cdot \tau_{\beta + \delta_{j}} = 0
\endaligned \tag{2.3.7}$$
for all $\alpha ,\beta \in L$ such that $(\alpha |\delta ) = -(\beta
|\delta ) = 1$.  Hirota's dot notation used here and further is
explained in Introduction (see (0.1.9)).

Equation (2.3.7) is known (see [DJKM1,2], [JM]) as the
$n$-component KP hierarchy of Hirota bilinear equations.  This equation still
describes the group orbit: $\sigma ({\Cal O}) = $ \newline
$\sigma R \sigma^{-1} (GL_{\infty}) \cdot 1$.

\vskip 10pt
\noindent
REMARK 2.3.  Equation (2.3.7) is invariant under the transformations
$\alpha\! \longmapsto\! \alpha\! +\! \gamma ,\ \beta \longmapsto
\beta + \gamma$,
where $\gamma \in M$.  Transformations of this type are called Schlessinger
transformations.

Let $\tau = \sum_{\gamma \in L} \tau_{\gamma}(x)e^{\gamma} \in B$;
the set $\text{supp}\ \tau
\overset{\text{def}}\to{=} \{ \gamma \in L|\ \tau_{\gamma} \neq 0\}$ is
called the {\it support} of $\tau$.

\proclaim{Proposition 2.3}  Let $\tau \in {\Bbb C}[[x]] \otimes {\Bbb
C}[M]$ be a solution to the KP hierarchy (2.3.4).  Then $\text{supp}\
\tau$ is a convex polyhedron with vertices in $M$ and edges parallel
to elements of $\Delta$.  More explicitly, $\text{supp}\ \tau$ may be
obtained by taking all possible sums of elements of a finite set
$S = \{ \gamma_{1} ,\ldots ,\gamma_{m}\}$, where all $\gamma_{i} \in \Delta$.
\endproclaim

\demo{Proof}  Consider the linear map ${\overline \sigma}:\ {\tilde
L} @>>> L$ defined by ${\overline \sigma}(\delta_{j}) = \delta_{(j +
1/2)\mod n}$, where\newline $a\mod n$ stands for the element of the set $\{
1,\ldots ,n\}$ congruent to $a\mod n$.  Then it is easy to see that
for $\tau \in F$ we have:

$$\text{supp}\ \sigma (\tau) = {\overline \sigma}(f\text{supp}\
\tau ).$$
Now Proposition 2.3 follows from Proposition 1.4.\ \ \ $\square$
\enddemo

\vskip 10pt

{\bf 2.4.}  The indeterminates $y^{(j)}_{k}$ in (2.3.7) are free parameters,
hence the coefficient of a monomial $y^{(j_{1})}_{k_{1}} \ldots
y^{(j_{s})}_{k_{s}}\ (k_{i} \in {\Bbb N},\ k_{1} \leq k_{2} \leq \ldots ,\
j_{i} \in \{ 1,\ldots ,n \} )$ in equation (2.3.7) gives us a Hirota bilinear
equation of the form

$$\sum^{n}_{i=1} \ \sum_{\alpha ,\beta} Q^{(j)}_{k;\alpha ,\beta} (D)
\tau_{\alpha -\delta_{i}} \cdot \tau_{\beta + \delta_{i}} = 0,
\tag{2.4.1}$$
where $Q^{(j)}_{k,\alpha ,\beta}$ are polynomials in the
$D^{(i)}_{r}$, $k = (k_{1},\ldots ,k_{s}),\ j = (j_{1},\ldots
,j_{s})$ and $\alpha
,\beta \in L$ are such that $(\alpha | \delta ) = - (\beta | \delta
) = 1$.  Each of these equations is a PDE in the indeterminates $x^{(j)}_{k}$
on functions $\tau_{\gamma},\ \gamma \in M$.

Recall that an expression $Q(D)\tau_{\alpha} \cdot \tau_{\beta}$ is
identically zero if and only if $\alpha = \beta$ and $Q(D) = -Q(-D)$.  The
corresponding Hirota bilinear equation is then called {\it trivial} and can be
disregarded.

Let us point out now that the energy decomposition (2.2.7) induces the
following energy decomposition on the space of Hirota bilinear equations:

$$\text{energy}(Q^{(j)}_{k;\alpha ,\beta} (D)\tau_{\alpha -\delta_{i}}
\cdot \tau_{\beta + \delta_{i}}) = k_{1} + \ldots + k_{s} + {\ssize
\frac{1}{2}}
((\alpha | \alpha)+(\beta | \beta)) \tag{2.4.2}$$
It is clear that the energy of a nontrivial Hirota bilinear equation is at
least
$2$.

Below we list the Hirota bilinear equations of lowest energy for
each $n$.

$n = 1$.  In this case we may drop the superscript in $D^{(1)}_{k}$ and the
subscript in $\tau_{\alpha}$ (which is $0$).  Each monomial $y_{k_{1} \ldots}
y_{k_{s}}$ gives a Hirota bilinear equation of the form

$$Q_{k}(D)\tau \cdot \tau = 0$$
of energy $k_{1} + \ldots + k_{s} + 1$.  An easy calculation shows that the
lowest energy of a non-trivial equation is $4$, and that there is a
unique non-trivial
equation of energy $4$, the classical KP equation in the Hirota bilinear form:

$$(D^{4}_{1} - 4D_{1}D_{3} + 3D^{2}_{2})\tau \cdot \tau = 0 . \tag{2.4.3}$$

$n \geq 2$.  There is an equation of energy $2$ for each unordered pair
of distinct indices $i$ and $k$ (recall that $\alpha_{ik} =
\delta_{i}-\delta_{k}$ are roots):

$$D^{(i)}_{1} D^{(k)}_{1} \tau_{0} \cdot \tau_{0} = 2
\tau_{\alpha_{ik}} \tau_{\alpha_{ki}}. \tag{2.4.4}$$
Furthermore, for each ordered pair of distinct indices $i$ and $j$
there are three  equations of energy $3$:

$$\align
&(D^{(i)}_{2} + D^{(i)2}_{1}) \tau_{0} \cdot
\tau_{\alpha_{ij}} = 0 , \tag{2.4.5} \\
&(D^{(j)}_{2} + D^{(j)2}_{1}) \tau_{\alpha_{ij}} \cdot
\tau_{0} = 0, \tag{2.4.6} \\
&D^{(i)}_{1} D^{(j)}_{2} \tau_{0} \cdot \tau_{0} + 2D^{(j)}_{1}
\tau_{\alpha_{ij}} \cdot
\tau_{\alpha_{ji}} = 0. \tag{2.4.7}
\endalign$$

$n \geq 3$.  There is an
equation of energy $2$ and an equation of energy $3$ for each ordered
triple of distinct indices
$i,j,k$:

$$D^{(k)}_{1} \tau_{0}\cdot \tau_{\alpha_{ij}} =
\varepsilon_{ik} \varepsilon_{kj} \varepsilon_{ij}
\tau_{\alpha_{ik}} \tau_{\alpha_{kj}},
 \tag{2.4.8}$$

$$D^{(k)}_{2} \tau_{0} \cdot \tau_{\alpha_{ij}} = \varepsilon_{ij}
\varepsilon_{kj} \varepsilon_{ik} D^{(k)}_{1} \tau_{\alpha_{ik}}
\cdot \tau_{\alpha_{kj}}. \tag{2.4.9}$$
(Note that (2.4.6) is a special case of (2.4.9) where $k = j$.)

$n \geq 4$.  There is an algebraic equation  of energy $2$ for each
ordered quadruple  of
distinct indices $i,j,k,\ell$:

$$\varepsilon_{ij}\varepsilon_{k\ell}\tau_{0}\tau_{\alpha_{ik}+\alpha_{j\ell}}
 + \varepsilon_{i\ell}\varepsilon_{jk}\tau_{\alpha_{ik}}
\tau_{\alpha_{j\ell}}
+ \varepsilon_{ik}
\varepsilon_{j\ell} \tau_{\alpha_{i\ell}}
\tau_{\alpha_{jk}} = 0 .  \tag{2.4.10} $$

Equations (2.4.4--10), together with an algebraic equation of energy
$3$ for each ordered sixtuple of distinct indices similar to
(2.4.10), form a complete list of non-trivial Hirota
bilinear equations of energy $\leq 3$ of the $n$-component KP hierarchy.

\vskip 10pt

\subheading{\S 3. The algebra of formal pseudo-differential operators and the
$n$-component KP hierarchy as a dynamical system}

\vskip 10pt
{\bf 3.0.} The KP hierarchy and its $n$-component generalizations admit
several  formulations.  The one given in the previous
section obtained by the field theoretical approach is the
$\tau$-function formulation given by Date, Jimbo, Kashiwara and Miwa
[DJKM1].  Another
well-known formulation, introduced by Sato [S], is given in the
language of formal pseudo-differential operators.  We will show that this
formulation follows from the $\tau$-function formulation given by
equation (2.3.3).

{\bf 3.1.} We shall work over the algebra ${\Cal A}$ of formal power
series over ${\Bbb C}$ in intederminates $x = (x^{(j)}_{k})$, where
$k = 1,2,\ldots $ and $j = 1,\ldots ,n$.  The indeterminates
$x^{(1)}_{1},\ldots ,x^{(n)}_{1}$ will be viewed as variables and
$x^{(j)}_{k}$ with $k \geq 2$ as parameters.  Let

$$\partial = \frac{\partial}{\partial x^{(1)}_{1}} + \ldots +
\frac{\partial}{\partial x^{(n)}_{1}}.$$
{\it A formal} $n \times n$ {\it matrix pseudo-differential
operator} is an expression of the form

$$P(x,\partial ) = \sum_{j \leq N} P_{j}(x)\partial^{j}, \tag{3.1.1}$$
where $P_{j}$ are $n \times n$ matrices over ${\Cal A}$.  The largest $N$
such that $P_{N} \neq 0$ is called the {\it order} of $P(x,\partial)$
(write $ord \ P(x,\partial ) = N$).  Let $\Psi$ denote the vector
space over ${\Bbb C}$ of all expressions (3.1.1).  We have a linear
isomorphism $s :\ \Psi @>>> \text{Mat}_{n}({\Cal A}((z)))$ given by
$s(P(x,\partial )) = P(x,z)$.  The matrix series
$P(x,z)$ in inteterminates $x$ and $z$ is called the {\it symbol} of
$P(x,\partial )$.

Now we may define a product $\circ$ on $\Psi$ making it an associative algebra:

$$s(P \circ Q)^{\hat{\ }} = \sum^{\infty}_{n = 0} \frac{1}{n!}
\frac{\partial^{n}s(P)}{\partial z^{n}} \partial^{n}s(Q).$$
We shall often drop the multiplication sign $\circ$ when no ambiguity
may arise.
Letting $\Psi (m) = \{ P \in \Psi |ord \ \Psi \leq m\}$, we get a
${\Bbb Z}$-filtration of the algebra $\Psi$:

$$\cdots \Psi (m+1) \supset \Psi (m) \supset \Psi (m-1) \supset
\cdots \tag{3.1.2}$$

One defines the differential part of $P(x,\partial )$ by
$P_{+}(x,\partial ) = \sum^{N}_{j=0} P_{j}(x) \partial^{j},$
and let $P_{-} = P-P_{+}$.  We have the corresponding vector space
decomposition:

$$\Psi = \Psi_{-} \oplus \Psi_{+}. \tag{3.1.3}$$

One defines a linear map $*: \Psi \rightarrow \Psi$ by the
following formula:
$$(\sum_{j} P_{j}\partial^{j})^{*} = \sum_{j} (-\partial )^{j}
\circ ^{t}P_{j}. \tag{3.1.4}$$
Here and further $^{t}P$ stands for the transpose of the matrix $P$.
Note that $*$ is an anti-involution of the algebra
$\Psi$.
In terms of symbols the anti-involution $*$ can be written in the
following closed form:

$$P^{*}(x,z) = (\exp \partial
\frac{\partial}{\partial z}) \ ^{t}P(x,-z). \tag{3.1.5}$$
It is clear that the anti-involution $*$ preserves the filtration
(3.1.2) and the decomposition (3.1.3).

\vskip 10pt
{\bf 3.2.}  Introduce the following notation

$$z \cdot x^{(j)} = \sum^{\infty}_{k=1} x^{(j)}_{k} z^{k},\ e^{z
\cdot x} = diag (e^{z\cdot x^{(1)}} ,\ldots ,e^{z\cdot x^{(n)}} ).$$
The algebra $\Psi$ acts on the space $U_{+}$ (resp.
$U_{-}$) of formal oscillating matrix functions of the form

$$\sum_{j \leq N} P_{j}z^{j}e^{z\cdot x}\ \ \ (\text{resp.}\ \sum_{j
\leq N} P_{j}z^{j} e^{-z\cdot x}), \ \text{where}\ \ P_{j} \in \
\text{Mat}_{n} ( {\Cal A} ),$$
in the obvious way:

$$P(x)\partial^{j}e^{\pm z\cdot x} = P(x)(\pm z)^{j} e^{\pm z\cdot x}.$$
We can now prove the following fundamental lemma.

\proclaim{Lemma 3.2} If $P,Q \in \Psi$ are such that

$$Res_{z=0} (P(x,\partial) e^{z\cdot x})  \ ^{t}
(Q(x^{\prime},\partial^{\prime}) e^{-z\cdot x^{\prime}})  dz = 0,
\tag{3.2.1}$$
then $(P \circ Q^{*})_{-} = 0$.
\endproclaim

\demo{Proof} Equation (3.2.1) is equivalent to

$$Res_{z=0} P(x,z)
e^{z(x-x^{\prime})}  \ ^{t}Q(x^{\prime},-z)  dz = 0. \tag{3.2.2}$$
The $(i,m)$-th entry of the matrix equation (3.2.2) is

$$Res_{z=0} \sum^{n}_{i=1} P_{ij}(x,z) Q_{mj}(x^{\prime},-z)e^{z\cdot
(x^{(j)}-x^{\prime (j)})}dz = 0.$$
Letting $y^{(j)}_{k} = x^{(j)}_{k} - x^{\prime (j)}_{k}$, this
equation can be rewritten by applying Taylor's formula to $Q$:

$$Res_{z=0} \sum^{n}_{j=1} P_{ij} (x,z) \exp \sum^{n}_{\ell =1}
\sum^{\infty}_{k=1} y^{(\ell)}_{k} (\delta_{\ell j} z^{k} -
\frac{\partial}{\partial x^{(\ell)}_{k}}) Q_{mj}(x,-z)dz = 0.
\tag{3.2.3}$$
Letting $y^{(\ell)}_{k} = 0$ for $k > 1$ and $y^{(\ell)}_{1} = y$ for
all $\ell$, we obtain from (3.2.3):

$$Res_{z=0} P(x,z) \sum_{k \geq 0} \frac{(-1)^{k}}{k!}
\partial^{k} (\ ^{t} Q) (x,-z)y^{k}e^{yz}dz = 0.
\tag{3.2.4}$$
Notice that $y^{k}e^{yz} = (e^{yz})^{(k)}$.  Here and further we
write $\varphi^{(k)}$ for $\frac{\partial^{k}\varphi}{\partial
z^{k}}$.  Using integration by parts with respect to $z$, (3.2.4)
becomes:

$$Res_{z=0} \sum_{k \geq 0} \frac{1}{k!} (  P(x,z)
\partial^{k}( \ ^{t}Q )(x,-z))^{(k)} e^{yz} dz = 0.
\tag{3.2.5}$$
Using Leibnitz formula, the left-hand side of (3.2.5) is equal to

$$\align
&Res_{z=0} \sum_{k \geq 0} \sum^{k}_{\ell = 0}
\frac{(-1)^{k-\ell}}{\ell ! (k-\ell )!}  P^{(\ell )} (x,z)
\left( \partial^{k}(\ ^{t}Q) \right)^{(k-\ell)} (x,-z) e^{yz} dz  \\
&  = Res_{z=0} \sum_{\ell \geq 0}
\frac{1}{\ell !}  P^{(\ell)}
(x,z) \partial^{\ell} \left(
\sum^{\infty}_{k=0} \frac{(-1)^{k}}{k!} \partial^{k} (\ ^{t}Q^{(k)})
(x,-z)\right)  e^{yz}dz \\
&  = Res_{z=0} \sum_{\ell \geq 0} \frac{1}{\ell !}  P^{(\ell
)}(x,z) \left( \frac{\partial^{\ell}  Q^{*}
(x,z)}{\partial x^{\ell}}\right) e^{yz} dz
= Res_{z=0} (P \circ Q^{*})(x,z) e^{yz} dz.
\endalign
$$

So we obtain that

$$Res_{z=0} ( P\circ  Q^{*})(x,z)e^{yz}dz = 0.
\tag{3.2.6}$$
Now write $(P \circ Q^{*})(x,z) = \sum_{j} A_{j}(x)
z^{j}$ and $e^{yz} = \sum^{\infty}_{\ell = 0}
\frac{(zy)^{\ell}}{\ell !}$.  Then from (3.2.6) we deduce:

$$0 = Res_{z=0} \sum_{j} \sum^{\infty}_{\ell = 0} A_{j}(x)
\frac{y^{\ell}}{\ell !} z^{\ell + j} dz = \sum^{\infty}_{\ell = 0}
A_{-\ell -1}(x) \frac{y^{\ell}}{\ell !}.$$
Hence $A_{j}(x) = 0$ for $j < 0$, i.e. $(P \circ Q^{*})_{-} =
0$.\ \ \ \ $\square$
\enddemo

\vskip 10pt
{\bf 3.3.}  We proceed now to rewrite the formulation (2.3.3) of the
$n$-component KP hierarchy in terms of formal pseudo-differential
operators.

Let $1 \leq a,b \leq n$ and recall formula (2.3.3) where $\alpha$ is
replaced by $\alpha + \delta_{a}$ and $\beta$ by $\beta -
\delta_{b}$:

$$\aligned
&\text{Res}_{z=0} (dz \sum^{n}_{j=1} \varepsilon (\delta_{j}, \alpha +
\delta_{a} - \beta + \delta_{b})
 z^{(\delta_{j}|\alpha + \delta_{a} -
\beta + \delta_{b} -2\delta_{j} )} \\
&\times \exp (\sum^{\infty}_{k=1} (x^{(j)^{\prime}}_{k} - x^{(j)^{\prime
\prime}}_{k})z^{k}) \exp(-\sum^{\infty}_{k=1}
(\frac{\partial}{\partial
x^{(j)^{\prime}}_{k}}-\frac{\partial}{\partial
x^{(j)^{\prime\prime}}_{k}}) \frac{z^{-k}}{k}) \\
&\tau_{\alpha + \alpha_{a_{j}}} (x^{\prime})
\tau_{\beta - \alpha_{b_{j}}}(x^{\prime\prime})) =
0\ \ \ \ (\alpha ,\beta \in M).
\endaligned \tag{3.3.1}$$

For each $\alpha \in \ \text{supp}\ \tau$ we define the (matrix
valued) functions
$$V^{\pm} (\alpha ,x,z) = (V^{\pm}_{ij}(\alpha ,x,z))^{n}_{i,j=1}
\tag{3.3.2}$$
as follows:

$$\aligned
&V^{\pm}_{ij}(\alpha ,x,z) \overset{\text{def}}\to{=}
\varepsilon (\delta_{j} , \alpha + \delta_{i})
 z^{(\delta_{j}|\pm \alpha + \alpha_{ij})} \\
& \times \exp (\pm \sum^{\infty}_{k=1} x^{(j)}_{k} z^{k})
\exp(\mp \sum^{\infty}_{k=1} \frac{\partial}{\partial
x^{(j)}_{k}} \frac{z^{-k}}{k}) \tau_{\alpha  \pm
\alpha_{ij}}  (x)/\tau_{\alpha}(x) .
\endaligned \tag{3.3.3}
$$
It is easy to see that equation (3.3.1) is equivalent to the
following bilinear identity:

$$Res_{z=0}V^{+}(\alpha ,x,z)\ ^{t}V^{-}(\beta ,x^{\prime},z)dz = 0\
\text{for all}\ \alpha ,\beta \in M. \tag{3.3.4}$$

Define $n \times n$ matrices $W^{\pm (m)} (\alpha ,x)$ by the
following generating series (cf. (3.3.3)):

$$
\sum^{\infty}_{m=0}
W^{\pm (m)}_{ij} (\alpha ,x)(\pm z)^{-m}
= \varepsilon_{ji}z^{\delta_{ij}-1} (\exp \mp
\sum^{\infty}_{k=1} \frac{\partial}{\partial x^{(j)}_{k}}\frac{z^{-k}}{k})
\tau_{\alpha \pm
\alpha_{ij}} (x))/\tau_{\alpha} (x). \tag{3.3.5}
$$
Note that

$$
W^{\pm (0)}(\alpha ,x) = I_{n}, \tag{3.3.6}$$
$$W^{\pm (1)}_{ij}(\alpha ,x) = \cases \varepsilon_{ji}
\tau_{\alpha \pm \alpha_{ij}}/\tau_{\alpha} &\text{if}\ i \neq j \\
- \tau^{-1}_{\alpha} \frac{\partial \tau_{\alpha}}{\partial
x^{(i)}_{1}} &\text{if}\ i = j, \endcases
\tag{3.3.7}$$

$$W^{\pm (2)}_{ij}(\alpha ,x) = \cases \mp \varepsilon_{ji}
\frac{\partial \tau_{\alpha \pm
\alpha_{ij}}} {\partial
x^{(j)}_{1}} /\tau_{\alpha} &\text{if}\ i \neq j, \\
(\mp \frac{1}{2} \frac{\partial \tau_{\alpha}}{\partial
x^{(i)}_{2}} + \frac{1}{2} \frac{\partial^{2} \tau_{\alpha}}{\partial
x^{(i)2}_{1}})/\tau_{\alpha} &\text{if}\ i = j.\endcases
\tag{3.3.8}$$

We see from (3.3.3) that $V^{\pm}(\alpha ,x,z)$ can be written in the
following form:

$$V^{\pm}(\alpha ,x,z) = (\sum^{\infty}_{m=0}
W^{\pm (m)}(\alpha ,x)R^{\pm}(\alpha ,\pm z)(\pm
z)^{-m})e^{\pm z \cdot x}, \tag{3.3.9}$$
where

$$R^{\pm}(\alpha ,z) = \sum^{n}_{i=1}
\varepsilon (\delta_{i}, \alpha ) E_{ii} (\pm z)^{\pm
(\delta_{i}|\alpha )}. \tag{3.3.10}$$
Here and further $E_{ij}$ stands for the $n \times n$ matrix whose
$(i,j)$ entry is $1$ and all other entries are zero.  Now it is clear
that $V^{\pm}(\alpha ,x,z)$ can be
written in
terms of formal pseudo-differential operators

$$P^{\pm}(\alpha ) \equiv P^{\pm} (\alpha ,x,\partial ) =
I_{n} + \sum^{\infty}_{m=1} W^{\pm (m)} (\alpha ,x)\partial^{-m}\
\text{and}\ R^{\pm}(\alpha ) = R^{\pm}(\alpha ,\partial) \tag{3.3.11}$$
as follows:
$$V^{\pm}(\alpha ,x,z) = P^{\pm } (\alpha )
R^{\pm}(\alpha)e^{\pm z \cdot x} . \tag{3.3.12}$$
Since obviously
$$R^{-}(\alpha ,\partial)^{-1} = R^{+}(\alpha
,\partial)^{*}, \tag{3.3.13}$$
using Lemma 3.2 we deduce from the bilinear identity
(3.3.4):

$$(P^{+}(\alpha)R^{+}(\alpha -\beta)P^{-}(\beta)^{*})_{-} = 0\
\text{for any}\ \alpha ,\beta \in \text{supp}\ \tau . \tag{3.3.14}$$
Furthermore, (3.3.14) for $\alpha = \beta$ is equivalent to

$$
P^{-}(\alpha  ) = (P^{+}(\alpha  )^{*})^{-1} ,
\tag{3.3.15}
$$
since $R^{\pm}(0) = I_{n}$ and $P^{\pm}(\alpha) \in I_{n} +
\Psi_{-}$.  Equations (3.3.14 and 15) imply

$$(P^{+}(\alpha)R^{+}(\alpha - \beta)P^{+}(\beta)^{-1})_{-} = 0\
\text{for all}\ \alpha ,\beta \in \text{supp}\ \tau . \tag{3.3.16}$$

{}From now on we shall write $P(\alpha)$ instead of $P^{+}(\alpha)$.

\proclaim{Proposition 3.3.}  Given $\beta \in \text{supp}\ \tau$, all the
pseudo-differential operators $P(\alpha )$, $\alpha \in \text{supp}\
\tau$, are
completely determined by $P(\beta )$ from equations (3.3.16).
\endproclaim

\demo{Proof}  We have for $i \neq j:\
R(\alpha_{ij}) = A\partial + B + C\partial^{-1}$,
where

$$A = \varepsilon_{ij} E_{ii},\ B = \sum^{n}_{\Sb k = 1\\ k \neq
i,j\endSb} \varepsilon_{ik}  \varepsilon_{jk} E_{kk},\ C =
\varepsilon_{ji} E_{jj}. \tag{3.3.17}$$
For $P = I_{n} + \sum^{\infty}_{j=1} W^{(j)}
\partial^{-j}$ we have

$$P^{-1} = I_{n} - W^{(1)}\partial^{-1} +
(W^{(1)2} - W^{(2)})\partial^{-2} + \cdots .
\tag{3.3.18}$$
Let $\alpha ,\beta \in M$ be such that $\alpha - \beta =
\alpha_{ij}$.  It follows from (3.3.18) and
(3.3.16) that $P(\alpha )R(\alpha -\beta)P(\beta)^{-1} =
(P(\alpha )R(\alpha -\beta)P(\beta )^{-1})_{+} = A\partial + B +
W^{(1)}(\alpha )A - AW^{(1)}(\beta),$ or equivalently:

$$P(\alpha ) (A\partial + B + C\partial^{-1}) = (A\partial + B +
W^{(1)}(\alpha )A - AW^{(1)}(\beta ))P(\beta).$$
Equating coefficients of $\partial^{-m}$, $m \geq 1$, we obtain:
$$\align
&W^{(m+1)}(\alpha)A + W^{(m)}(\alpha)B + W^{(m-1)}(\alpha )C = \\
&= A(\partial W^{(m)}(\beta )W^{(m+1)}(\beta ) - W^{(1)}(\beta
)W^{(m)}(\beta )) + BW^{(m)}(\beta ) + W^{(1)}(\alpha )AW^{(m)}(\beta
).
\endalign$$
Substituting expressions (3.3.17) for $A,\ B$ and $C$, we obtain an
explicit form of matching conditions
$(m \geq 1)$:

$$\aligned
&\varepsilon_{ij}W^{(m+1)}(\alpha)E_{ii} + \sum_{k \neq i,j}
\varepsilon_{ik} \varepsilon_{jk} W^{(m)}(\alpha)E_{kk} +
\varepsilon_{ji}W^{(m-1)}(\alpha) E_{jj} \\
&= \varepsilon_{ij} E_{ii} (\partial W^{(m)}(\beta ) +
W^{(m+1)}(\beta ) - W^{(1)} (\beta ) W^{(m)}(\beta)) \\
&+ \sum_{k \neq i,j} \varepsilon_{ik} \varepsilon_{jk}
E_{kk}W^{(m)}(\beta ) + \varepsilon_{ij} W^{(1)}(\alpha ) E_{ii}
W^{(m)}(\beta).
\endaligned \tag{3.3.19}
$$
It follows from (3.3.19) that $W^{(m+1)}(\alpha)$ for $m \geq 1$ can
be expressed in terms of the $W^{(s)}(\beta )$ with $s \leq m+1$ and
$W^{(1)}(\alpha)$.  Looking at the $(k,\ell)$-entry of (3.3.19) for
$k,\ell \neq i,j$, we see that $W^{(1)}(\alpha)$ can be expressed in
terms of $W^{(1)}(\beta)$ and $W^{(1)}_{ki}(\alpha)$, where $k \neq
i,j$.  The $(k,j)$-entry of (3.3.19) for $m = 1$ gives:
$W^{(1)}_{ki}(\alpha) W^{(1)}_{ij}(\beta) =
\varepsilon_{ik}\varepsilon_{jk} W^{(1)}_{kj}(\beta)$, and since the
$(j,j)$-entry of this equation is
$W^{(1)}_{ji}(\alpha)W^{(1)}_{ij}(\beta) = -1$, we see that
$W^{(1)}_{ij}(\beta)$ is invertible, hence $W^{(1)}_{ki}(\alpha)$ is
expressed in terms of $W^{(1)}(\beta)$.

Due to Proposition 2.4 for any $\alpha , \beta \in \ \text{supp}\
\tau$ there exist a sequence $\gamma_{1},\ldots ,\gamma_{k}$ such
that $\alpha = \gamma_{1},\ \beta = \gamma_{k}$ and $\gamma_{i} -
\gamma_{i+1}  \in \Delta$ for all $i = 1,\ldots ,k-1$.  The
proposition now follows. \ \ \ $\square$
\enddemo

\vskip 10pt
\noindent
REMARK 3.3.  The functions $P^{+}(\alpha ,x,z)\ (\alpha \in M)$
determine the
$\tau$-function \newline
$\sum_{\alpha}$ $\tau_{\alpha}(x)e^{\alpha}$ up to a constant factor.  Namely,
we may recover $\tau_{\alpha}(x)$ from functions $P^{+}_{jj}(\alpha
,x,z)$ as follows.  We have from (3.3.5):

$$\log P^{+}_{jj}(\alpha ,x,z) = \log \tau_{\alpha} (x^{(p)}_{\ell} -
\frac{\delta_{jp}}{\ell z^{\ell}}) - \log \tau_{\alpha}
(x^{(p)}_{\ell}).$$
Applying to both sides the operator $\frac{\partial}{\partial z} -
\sum_{k\geq 1} z^{-k-1} \frac{\partial}{\partial x^{(j)}_{k}}$ (that
kills the first summand on the right), we obtain:

$$
(\frac{\partial}{\partial z} - \sum_{k \geq 1} z^{-k-1}
\frac{\partial}{\partial x^{(j)}_{k}}) \log P^{+}_{jj}(\alpha ,x,z) =
 \sum_{k \geq 1} z^{-k-1} \frac{\partial}{\partial x^{(j)}_{k}} \log
\tau_{\alpha}(x).
$$
Hence
$$\frac{\partial}{\partial x^{(j)}_{k}} \log \tau_{\alpha}(x) =
Res_{z=0}\  dz\  z^{k} (\frac{\partial}{\partial z} - \sum_{k \geq 1}
z^{-k-1} \frac{\partial}{\partial  x^{(j)}_{k}}) \log P^{+}_{jj}(\alpha
,x,z). \tag{3.3.20}$$
This determines $\tau_{\alpha}(x)$ up to a constant factor.  It
follows from (3.3.7) and Proposition 2.3 that these constant factors
are the same for
all $\alpha$.

\vskip 10pt
{\bf 3.4.}  Introduce the following formal pseudo-differential
operators $L(\alpha),\  C^{(j)}(\alpha ),\ L^{(j)}(\alpha)$ and
differential operators $B_{m}(\alpha )$ and $B^{(j)}_{m}(\alpha)$:

$$\aligned
L(\alpha ) \equiv L(\alpha ,x,\partial)
  & = P^{+}(\alpha) \circ \partial \circ P^{+}(\alpha)^{-1}, \\
C^{(j)}(\alpha) \equiv C^{(j)}(\alpha ,x,\partial) &=
P^{+}(\alpha)E_{jj} P^{+}(\alpha)^{-1}, \\
L^{(j)}(\alpha) \equiv C^{(j)}(\alpha)L(\alpha)  &= P^{+}(\alpha)E_{jj}
\circ \partial \circ P^{+}(\alpha)^{-1}, \\
B_{m}(\alpha) \equiv (L(\alpha)^{m})_{+} &= (P^{+}(\alpha)\circ
\partial^{m} \circ P^{+}(\alpha)^{-1})_{+}, \\
B^{(j)}_{m}(\alpha) \equiv (L^{(j)}(\alpha)^{m})_{+} &=
(P^{+}(\alpha)E_{jj} \circ \partial^{m} \circ
P^{+}(\alpha)^{-1})_{+}.
\endaligned \tag{3.4.1}
$$
Using Lemma 3.2 we can now derive the Sato equations from equation
(3.3.4):

\proclaim{Lemma 3.4}  Each formal pseudo-differential operator $P =
P^{+}(\alpha)$ satisfies the Sato equations:

$$
\frac{\partial P}{\partial x^{(j)}_{k}} = -(PE_{jj} \circ
\partial^{k} \circ P^{-1})_{-} \circ P. \tag{3.4.2}$$

\endproclaim

\demo{Proof} Notice first that
$$\align
&(\frac{\partial}{\partial x^{(j)}_{k}}-B^{(j)}_{k}(\alpha))
V^{+}(\alpha ,x,z)  = (\frac{\partial}{\partial x^{(j)}_{k}} -
B^{(j)}_{k}(\alpha))P^{+}(\alpha) R^{+}(\alpha) e^{z\cdot x} \\
&= (\frac{\partial P^{+}(\alpha)}{\partial x^{(j)}_{k}} R^{+}(\alpha)
+ P^{+}(\alpha) R^{+}(\alpha ) E_{jj} \partial^{k} -
B^{(j)}_{k}(\alpha)P^{+}(\alpha)R^{+}(\alpha))e^{z\cdot x} \\
&= (\frac{\partial P^{+}(\alpha)}{\partial x^{(j)}_{k}} + P^{+}(\alpha)
E_{jj} \partial^{k} - B^{(j)}_{k}(\alpha)P^{+}(\alpha))R^{+}(\alpha)
e^{z\cdot x} \\
&= (\frac{\partial P^{+}(\alpha)}{\partial x^{(j)}_{k}} +
L^{(j)}(\alpha)^{k} P^{+}(\alpha) - B^{(j)}_{k}(\alpha)
P^{+}(\alpha))R^{+}(\alpha) e^{z\cdot x} \\
&= (\frac{\partial P^{+}(\alpha)}{\partial x^{(j)}_{k}} +
(L^{(j)}(\alpha)^{k})_{-} P^{+}(\alpha)) R^{+}(\alpha) e^{z \cdot x}
\endalign$$
Next apply $\frac{\partial}{\partial
x^{(j)}_{k}}-B^{(j)}_{k}(\alpha)$ to the equation (3.3.4) for $\alpha
= \beta$ to obtain:
$$Res_{z=0}\ dz \ (\frac{\partial P^{+}(\alpha)}{\partial x^{(j)}_{k}} +
(L^{(j)}(\alpha)^{k})_{-})(P^{+}(\alpha)R^{+}(\alpha)e^{z\cdot
x})\ ^{t}(P^{-}(\alpha)R^{-}(\alpha) e^{-z\cdot x^{\prime}}) = 0.$$
Now apply Lemma 3.2 and (3.3.15) to obtain:

$$((\frac{\partial P^{+}(\alpha)}{\partial x^{(j)}_{k}} +
(L^{(j)}(\alpha)^{k})_{-} P^{+}(\alpha))P^{+}(\alpha)^{-1})_{-} = 0$$
which proves the lemma.\ \ \ \ \ \ \ \ $\square$
\enddemo

\proclaim{Proposition 3.4}  Consider the formal oscillating functions
$V^{+}(\alpha ,x,z)$ and $V^{-}(\alpha ,x,z),$ \newline $\alpha \in M$, of the
form (3.3.12), where $R^{\pm}(\alpha ,z)$ are given by (3.3.10) and
$P^{\pm}(\alpha ,x,\partial ) \in I_{n} + \Psi_{-}$.  Then the
bilinear identity (3.3.4) for all $\alpha ,\beta \in \ \text{supp}\
\tau$ is equivalent to the Sato equation (3.4.2) for each $P =
P^{+}(\alpha )$ and the matching condition (3.3.14) for
all $\alpha ,\beta \in \ \text{supp} \ \tau$.
\endproclaim

\demo{Proof} We have proved already that the bilinear identity
(3.3.4) implies (3.4.2) and (3.3.14).  To prove the converse, denote
by $A(\alpha ,\beta ,x,x^{\prime})$ the left-hand side of (3.3.4).
The same argument as in the proof of Lemma 3.4 shows that:

$$\left( \frac{\partial}{\partial x^{(j)}_{k}} - B^{(j)}_{k}(\alpha
)\right) A(\alpha ,\beta ,x,x^{\prime}) = 0 , \tag{3.4.3}$$

$$A(\alpha ,\beta ,x,x) = 0 , \tag{3.4.4}$$
where $B^{(j)}_{k}(\alpha )$ is defined by (3.4.1).

Denote by $A_{1}(\alpha ,\beta )$ the expression for $A (\alpha
,\beta ,x,x^{\prime})$ in which we set $x^{(j)}_{k} = x^{\prime
(j)}_{k} = 0$ if $k \geq 2$ and $x^{(1)}_{1} = \ldots = x^{(n)}_{1} =
x_{1},\ x^{\prime (1)}_{1} = \ldots = x^{\prime (n)}_{1} =
x^{\prime}_{1}$.  Expanding $A(\alpha ,\beta ,x,x^{\prime})$ in a
power series with coefficients in ${\Bbb C}[[x^{(1)}_{1} + \ldots +
x^{(n)}_{1}]]$, we deduce from (3.4.3) that all its coefficients are
multiples of the constant term.  Hence, due to (3.4.4) it remains to
prove that

$$A_{1}(\alpha ,\beta ) = 0. \tag{3.4.5}$$
But the same argument as in the proof of Lemma 3.2 shows that

$$A_{1} (\alpha ,\beta ) = \text{Res}_{z=0} P^{+}(\alpha
,x_{1},\partial ) R^{+}(\alpha -\beta ,\partial ) P^{-}(\beta
,x_{1},\partial )^{*}\ e^{yz}dz ,$$
where $y = x_{1} - x^{\prime}_{1}$.  Hence, as at the end of the
proof of Lemma 3.2, (3.4.5) follows from (3.3.14).\ \ \ $\square$
\enddemo

\vskip 10pt
{\bf 3.5.}  Fix $\alpha \in M$; we have introduced above a collection of
formal pseudo-differential operators $L \equiv L(\alpha),\ C^{(i)} \equiv
C^{(i)}(\alpha)$ of the form:

$$\aligned
L &= I_{n} \partial + \sum^{\infty}_{j=1} U^{(j)}(x) \partial^{-j} , \\
C^{(i)} &= E_{ii} + \sum^{\infty}_{j=1} C^{(i,j)}(x)
\partial^{-j},\ i = 1,2,\cdots ,n,
\endaligned \tag{3.5.1}$$
subject to the conditions

$$
\sum^{n}_{i=1} C^{(i)} = I_{n}, \
C^{(i)}L = LC^{(i)},\ C^{(i)}C^{(j)} = \delta_{ij} C^{(i)}.
 \tag{3.5.2}$$
They satisfy the following set of equations for some $P \in I_{n} + \Psi_{-}$:

$$
\cases LP = P\partial &\ \\
C^{(i)}P = PE_{ii} &\ \\
\frac{\partial P}{\partial x^{(i)}_{k}} = -(L^{(i)k})_{-} P,\
\text{where}\  L^{(i)} = C^{(i)}L.
\endcases
\tag{3.5.3}$$
Notice that the first equation of (3.5.3) follows from the last one,
since $L = I_{n}\partial + \sum_{i} (L^{(i)})_{-}$.

\proclaim{Proposition 3.5}  The system of equations (3.5.3) has a
solution $P \in I_{n} + \Psi_{-}$ if and only if we can find a formal
oscillating function of the form

$$W(x,z) = (I_{n} + \sum^{\infty}_{j=1} W^{(j)}(x)z^{-j})e^{z\cdot x}
\tag{3.5.4}$$
that satisfies the linear equations

$$
LW = zW ,\
C^{(i)}W = WE_{ii} , \
\frac{\partial W}{\partial x^{(i)}_{k}} = B^{(i)}_{k}W .
 \tag{3.5.5}$$
\endproclaim

Proof (3.5.3) $\Rightarrow$ (3.5.5):  Put $W = Pe^{z\cdot x}$.  Then
we have:

$$\align
LW &= LPe^{z\cdot x} = P\partial e^{z\cdot x} = zPe^{z\cdot x} = zW;
\\
C^{(i)}W &= C^{(i)}Pe^{z\cdot x} = PE_{ii}e^{z\cdot x} = Pe^{z\cdot
x}E_{ii} = WE_{ii}; \\
\frac{\partial W}{\partial x^{(i)}_{k}} &=
\frac{\partial P}{\partial x^{(i)}_{k}} + P \frac{\partial e^{z\cdot
x}}{\partial x^{(i)}_{k}} = -(L^{(i)k})_{-} Pe^{z\cdot x} + z^{k}
PE_{ii} e^{z\cdot x} \\
&= -(L^{(i)k})_{-} W + PE_{ii} \partial^{k} e^{z\cdot x} =
-(L^{(i)k})_{-}W + C^{(i)}P\partial^{k} e^{z\cdot x} \\
&= - (L^{(i)k})_{-} W + C^{(i)}L^{k}Pe^{z\cdot x} = -(L^{(i)k})_{-}W
+ L^{(i)k}W = B^{(i)}_{k}W.
\endalign$$

(3.5.5) $\Rightarrow$ (3.5.3).  Define $P \in \Psi$ by $W = Pe^{z
\cdot x}$.  If $LW = zW$, then $LPe^{z\cdot x} = zPe^{z\cdot x} =
P\partial e^{z\cdot x}$, hence $LP = P\partial$.

If $C^{(i)}W = WE_{ii}$, then $C^{(i)}Pe^{z\cdot x} = Pe^{z\cdot
x}E_{ii} = PE_{ii}e^{z\cdot x}$, hence $C^{(i)}P = PE_{ii}$.

Finally, the last equation of (3.5.5) gives:
$\frac{\partial}{\partial x^{(i)}_{k}} (Pe^{z \cdot x}) =
-(L^{(i)k})_{-} Pe^{z \cdot x} + L^{(i)k}Pe^{z \cdot x}$.  Since we
have already proved the first two equations of (3.5.3), we derive (as
above): $L^{(i)k}Pe^{z\cdot x} = z^{k}Pe^{z\cdot x} = P
\frac{\partial e^{z \cdot x}}{\partial x^{(i)}_{k}}$, hence:
$\frac{\partial P}{\partial x^{(i)}_{k}} e^{x\cdot z} =
-(L^{(i)k})_{-}P e^{x \cdot z}$, which proves that $P$ satisfies the
Sato equations.\ \ \ \ $\square$

\vskip 10pt
\noindent
REMARKS 3.5. (a) It is easy to see that the collection of formal
pseudo-differential operators $\{ L,C^{(1)},\ldots ,C^{(n)} \}$ of
the form (3.5.1)
and satisfying (3.5.2) can be simultaneously conjugated to the
trivial collection  $\{ \partial ,E_{11} ,\ldots , E_{nn} \}$ by some
$P \in I_{n} +
\Psi_{-}$.  It follows that the solution of the form (3.5.4) to the
linear problem (3.5.5) is unique up to multiplication on the right by
a diagonal matrix of the form

$$D(z) = \exp - \sum^{\infty}_{j=1} a_{j}z^{-j}/j , \tag{3.5.6}$$
where the $a_{j}$ are diagonal matrices over ${\Bbb C}$ (indeed, this
is the case for the trivial collection).  The space of all solutions
of (3.5.5) in formal oscillating functions is obtained from one of the
form (3.5.4) by multiplying on the right by a diagonal matrix over
${\Bbb C}((z))$.  For that reason the (matrix valued) functions

$$W^{+}(\alpha ,x,z) = P^{+}(\alpha)e^{x\cdot z},\ \alpha \in \
\text{supp}\ \tau , \tag{3.5.7}$$
are called the {\it wave functions} for $\tau$.
The formal pseudo-differential operator $P^{+}(\alpha )$ is called
the wave operator.  The
functions $W^{-}(\alpha ,x,z) = P^{-}(\alpha ) e^{-x\cdot z}$ are
called the adjoint wave functions
and the operators $P^{-}(\alpha )$
(which are expressed via $P^{+}(\alpha )$ by (3.3.15)) are called the
adjoint wave operators.  Note that $V^{+}(\alpha , x,z)$ are
solutions of (3.5.5) as well since they are obtained by multiplying
$W^{+}(\alpha ,x,z)$ on the right by $R^{+}(\alpha ,z)$.

(b) Multiplying the wave function $W^{+}(\alpha ,x,z)$ on the right
by $D(z)$ given by (3.5.6) corresponds to multiplying the
corresponding $\tau$-function by $\exp tr \sum^{\infty}_{k=1}
a_{k}x_{k}$, where $x_{k} = \text{diag}\ (x^{(1)}_{k},\ldots ,x^{(n)}_{k})$.

(c) The collection $\{ L,C^{(1)} , \ldots , C^{(n)} \}$ determines
uniquely $P \in I_{n}
+ \Psi_{-}$ up to the multiplication of $P$ on the right by a
formal pseudo-differential operator with constant coefficients from
$I_{n} + \Psi_{-}$.

\vskip 20pt
{\bf 3.6.}  In this section we shall rewrite the compatibility
conditions (3.5.3) (or equivalent compatibility conditions (3.5.5)) in
the form of Lax equations and Zakharov-Shabat equations.

\proclaim{Lemma 3.6} If for every $\alpha \in M$ the
formal pseudo-differential operators $L \equiv L(\alpha)$ and $C^{(j)} \equiv
C^{(j)}(\alpha)$ of the form (3.5.1) satisfy conditions (3.5.2) and if the
equations (3.5.3) have a solution $P \equiv P(\alpha) \in I_{n} +
\Psi_{-}$, then the
differential operators $B^{(j)}_{k} \equiv B^{(j)}_{k}(\alpha) =
(L^{(j)}(\alpha)^{k})_{+}$ satisfy one of the following equivalent
conditions:

$$ \left\{ \matrix \frac{\partial L}{\partial x^{(j)}_{k}} =
[B^{(j)}_{k},L]  & \   \\
\frac{\partial C^{(i)}}{\partial x^{(j)}_{k}} = [B^{(j)}_{k},C^{(i)}] & \
\  \endmatrix \right. \tag{3.6.1}$$

$$\frac{\partial L^{(i)}}{\partial x^{(j)}_{k}} =
[B^{(j)}_{k},L^{(i)}] \tag{3.6.2}$$

$$\frac{\partial B^{(i)}_{\ell}}{\partial x^{(j)}_{k}} -
\frac{\partial B^{(j)}_{k}}{\partial x^{(i)}_{\ell}} =
[B^{(j)}_{k},B^{(i)}_{\ell}]. \tag{3.6.3}$$

\endproclaim

\demo\nofrills{Proof} (cf. [Sh]): To derive the first equation of
(3.6.1) we differentiate
the equation $LP = P\partial$ by $x^{(j)}_{k}$:

$$\frac{\partial L}{\partial x^{(j)}_{k}} P + L \frac{\partial
P}{\partial x^{(j)}_{k}} = \frac{\partial P}{\partial x^{(j)}_{k}}
\partial,$$
and substitute Sato's equation (see (3.5.3)).  Then one obtains:

$$\frac{\partial L}{\partial x^{(j)}_{k}} P = (B^{(j)}_{k}L -
LB^{(j)}_{k})P$$
from which we derive the desired result.  The second equation of
(3.6.1) is proven analogously:\ differentiate $C^{(i)}P = PE_{ii}$,
substitute Sato's equation and use the fact that $[L^{(j)k},C^{(i)}]
= 0$.

Next we prove the equivalence of (3.6.1), (3.6.2) and (3.6.3).  The
implication (3.6.1) $\Rightarrow$ (3.6.2) is trivial.  To prove the
implication (3.6.2) $\Rightarrow$ (3.6.1) note that $L =
\sum^{n}_{j=1} L^{(j)}$ implies that the first equation of (3.6.1)
follows immediately.  As for the second one, we have:

$$\align
\frac{\partial C^{(i)}}{\partial x^{(j)}_{k}} &= (\frac{\partial
L^{(i)}}{\partial x^{(j)}_{k}} - C^{(i)} \frac{\partial L}{\partial
x^{(j)}_{k}})L^{-1} \\
&= ([B^{(j)}_{k},L^{(i)}] - C^{(i)}[B^{(j)}_{k},L])L^{-1} \\
&= ([B^{(j)}_{k},C^{(i)}]L)L^{-1} = [B^{(j)}_{k},C^{(i)}].
\endalign$$

Next, we prove the implication (3.6.2) $\Rightarrow$ (3.6.3).  Since
both $\frac{\partial}{\partial x^{(j)}_{k}}$ and $ad\ B^{(j)}_{k}$
are derivations, (3.6.2) implies:

$$\frac{\partial L^{(i)\ell}}{\partial x^{(j)}_{k}} =
[B^{(j)}_{k},L^{(i)\ell}].$$
Hence:
$$\align
&\left( \frac{\partial B^{(i)}_{\ell}}{\partial x^{(j)}_{k}} -
\frac{\partial B^{(j)}_{k}}{\partial x^{(i)}_{\ell}} -
[B^{(j)}_{k},B^{(i)}_{\ell}] \right) +
 \left( \frac{\partial (L^{(i)\ell})_{-}}{\partial x^{(j)}_{k}} -
\frac{\partial (L^{(j)k})_{-}}{\partial x^{(i)}_{\ell}} +
[(L^{(j)k})_{-},(L^{(i)\ell})_{-}] \right) \\
&= [B^{(j)}_{k},L^{(i)\ell}] -
[B^{(i)}_{\ell},L^{(j)k}]-[B^{(j)}_{k},B^{(i)}_{\ell}]
 + [(L^{(j)k})_{-},(L^{(i)\ell})_{-}] \\
&= [L^{(j)k},L^{(i)\ell}] = 0.
\endalign$$
Since $\Psi_{-} \cap \Psi_{+} = \{ 0 \}$, both terms on the
left-hand  side are zero proving (3.6.3).

Finally, we prove the implication (3.6.3) $\Rightarrow$ (3.6.2).  We
rewrite (3.6.3):

$$\frac{\partial L^{(i)\ell}}{\partial x^{(j)}_{k}} -
[B^{j}_{k},L^{(i)\ell}] = \frac{\partial (L^{(i)\ell})_{-}}{\partial
x^{(j)}_{k}} + \frac{\partial B^{(j)}_{k}}{\partial x^{(i)}_{\ell}} -
[B^{(j)}_{k},(L^{(i)\ell})_{-}]$$
This right-hand side has order $k-1$, hence

$$\frac{\partial L^{(i)\ell}}{\partial x^{(j)}_{k}} -
[B^{(j)}_{k},L^{(i)\ell}] \in \Psi (k-1)\ \text{for every}\ \ell > 0.
\tag{3.6.4}$$
Now suppose that $\frac{\partial L^{(i)}}{\partial x^{(j)}_{k}} -
[B^{(j)}_{k},L^{(i)}] \neq 0$.  Then:

$$\lim_{\ell \rightarrow \infty} ord(\frac{\partial
L^{(i)\ell}}{\partial x^{(i)}_{k}} - [B^{(j)}_{k},L^{(i)\ell}]) =
\infty$$
which contradicts (3.6.4).\ \ \ \ \ \ \ $\square$

\enddemo

Equations (3.6.1) and (3.6.2) are called {\it Lax type} equations.
Equations (3.6.3) are called the {\it Zakharov-Shabat type}
equations.  The latter
are the compatibility conditions for the linear problem (3.5.4).
Indeed, since $\frac{\partial}{\partial x^{(j)}_{k}}
\frac{\partial}{\partial x^{(i)}_{\ell}} \ W =
\frac{\partial}{\partial x^{(i)}_{\ell}} \frac{\partial}{\partial
x^{(j)}_{k}} W$, one finds

$$
0 = \frac{\partial}{\partial x^{(j)}_{k}} (B^{(i)}_{\ell}W) -
\frac{\partial}{\partial x^{(i)}_{\ell}} (B^{(j)}_{k}W)
= (\frac{\partial B^{(i)}_{\ell}}{\partial x^{(j)}_{k}}
 - \frac{\partial B^{(j)}_{k}}{\partial
x^{(i)}_{\ell}} - [B^{(j)}_{k}, B^{(i)}_{\ell}])W.
$$

Notice that as a byproduct of the proof of Proposition 3.6, we obtain
complementary Zakharov-Shabat equations:

$$\frac{\partial (L^{(i)\ell})_{-}}{\partial x^{(j)}_{k}} -
\frac{\partial (L^{(j)k})_{-}}{\partial x^{(i)}_{\ell}} =
[(L^{(i)\ell})_{-},(L^{(j)k})_{-}]. \tag{3.6.5}$$

\proclaim{Proposition 3.6} Sato equations (3.4.2) on $P \in I_{n} +
\Psi_{-}$ imply equations (3.6.3) on differential operators
$B^{(i)}_{k} = (L^{(i)k})_{+}$.
\endproclaim

\noindent
{\it Proof,} is the same as that of the corresponding part of Lemma
3.6.\ \ \ \ $\square$

\vskip 10pt
\noindent
REMARK 3.6.  The above results may be summarized as follows.
The $n$-component KP hierarchy (2.3.7) of Hirota bilinear equations
on the $\tau$-function is equivalent to the bilinear equation
(3.3.4) on the wave function, which is related to the
$\tau$-function by formula (3.3.3) and Remark 3.3.  The bilinear
equation (3.3.4) for each $\alpha = \beta$ implies the Sato equation
(3.4.2) on the formal pseudo-differential operator $P \equiv
P(\alpha)$.  Moreover, equation (3.4.2) on $P(\alpha )$ for each
$\alpha$ together with the matching conditions (3.3.14) are
equivalent to the bilinear identity (3.3.4).  Also, the Sato equation
(or rather (3.5.3)) is a compatibility condition for the linear
problem (3.5.5) for the wave function. The
Sato equation in turn implies the  system of Lax
type equations (3.6.2) (or equivalent systems (3.6.1) or (3.6.3),
which is the most familiar form of the compatibility condition) on
formal pseudo-differential operators $L^{(i)}$ (resp. $L$ and $C^{(i)}$
satisfying costrainsts
(3.5.2)).  The latter formal pseudo-differential operators are expressed via
the wave function by formulas (3.4.1), (3.3.9--12).

\vskip 10pt
{\bf 3.7.}  In this section we write down explicitly some of the Sato
equations (3.4.2) on the matrix elements $W^{(s)}_{ij}$ of the
coefficients $W^{(s)}(x)$ of the pseudo-differential operator

$$P = I_{n} + \sum^{\infty}_{m=1} W^{(m)}(x) \partial^{-m}.$$
We shall write $W_{ij}$ for $W^{(1)}_{ij}$ to simplify notation.  We
have for $i \neq k$:

$$\frac{\partial W_{ij}}{\partial x^{(k)}_{1}} = W_{ik} W_{kj} -
\delta_{jk} W^{(2)}_{ij}, \tag{3.7.1}$$

$$\frac{\partial W^{(2)}_{ij}}{\partial x^{(k)}_{1}} = W_{ik}
W^{(2)}_{kj} - \delta_{jk} W^{(3)}_{ij}. \tag{3.7.2}$$
Next, calculating $\frac{\partial W_{ij}}{\partial x^{(k)}_{2}}$ from
(3.4.2) and substituting (3.7.1) and (3.7.2) in these equations, we
obtain:

$$\frac{\partial W_{ij}}{\partial x^{(k)}_{2}} = W_{ik}
\frac{\partial W_{kj}}{\partial x^{(k)}_{1}} - \frac{\partial
W_{ik}}{\partial x^{(k)}_{1}} W_{kj} \ \text{if}\ k \neq i\
\text{and}\ k \neq j , \tag{3.7.3}$$

$$\frac{\partial W_{ij}}{\partial x^{(j)}_{2}} = 2 \frac{\partial
W_{jj}}{\partial x^{(j)}_{1}}  W_{ij} - \frac{\partial^{2}
W_{ij}}{\partial x^{(j)2}_{1}}\ \text{if}\ i \neq j, \tag{3.7.4}$$

$$\frac{\partial W_{ij}}{\partial x^{(i)}_{2}} = -2 \frac{\partial
W_{ii}}{\partial x^{(i)}_{1}} W_{ij} +
\frac{\partial^{2}W_{ij}}{\partial x^{(i)2}_{1}} \ \text{if}\ i \neq
j, \tag{3.7.5}$$

$$\frac{\partial W_{ii}}{\partial x^{(i)}_{2}} =
\frac{\partial^{2}W_{ii}}{\partial x^{(i)2}_{1}} + 2 \sum_{p \neq i}
W_{ip} \frac{\partial W_{pi}}{\partial x^{(i)}_{1}} - 2W_{ii}
\partial W_{ii} + 2\partial W^{(2)}_{ii}. \tag{3.7.6}$$

\noindent
REMARK 3.7.  Substituting expressions for the $W_{ij} = W^{(1)}_{ij}
(\alpha = 0,x)$ given by (3.3.8), the above equations
turn into the Hirota bilinear equations found in \S 2.4 as follows:

$$\matrix \format \l &\quad \c &\quad \l \\
(3.7.1)\ \text{for}\ i = j &\Rightarrow &(2.4.4)\\
(3.7.1)\ \text{for}\ i \neq j &\Rightarrow &(2.4.8)\\
(3.7.5)   &\Rightarrow &(2.4.5), \\
(3.7.4)   &\Rightarrow &(2.4.6), \\
(3.7.3)\ \text{for}\ i = j   &\Rightarrow &(2.4.7)\ (\text{with}\ j
\ \text{replaced by}\ k), \\
(3.7.3) \ \text{for}\ i \neq j &\Rightarrow &(2.4.9).
\endmatrix$$

\vskip 10pt
{\bf 3.8.}  In this section we write down explicitly some of the Lax
equations (3.6.1) of the $n$-component KP hierarchy and auxiliary
conditions (3.5.2) for the formal pseudo-differential operators

$$
L = I_{n}\partial + \sum^{\infty}_{j=1} U^{(j)}(x) \partial^{-j} \
\text{and}\
C^{(i)} = E_{ii} + \sum^{\infty}_{j=1} C^{(i,j)} (x) \partial^{-j} \
(i = 1,\ldots ,n).  \tag{3.8.1}$$
For the convenience of the reader, recall that $x$ stands for all
indeterminates $x^{(k)}_{i}$, where $i = 1,2,\ldots$ and $k =
1,\ldots ,n$, that the auxiliary conditions are

$$\sum^{n}_{i=1} C^{(i)} = I_{n},\ C^{(i)}C^{(j)} =
\delta_{ij}C^{(i)},\ C^{(i)}L = LC^{(i)}, \tag{3.8.2}$$
and that the Lax equations of the $n$-component KP hierarchy are

\noindent
(3.8.3a)$_{i}$\ \ \ \ \ \ \ \ \ \ \ $\frac{\partial L}{\partial x^{(k)}_{i}} =
[B^{(k)}_{i},L],$
\vskip 5pt
\noindent
(3.8.3b)$_{i}$\ \ \ \ \ \ \ $\frac{\partial C^{(\ell)}}{\partial
x^{(k)}_{i}} =
[B^{(k)}_{i},C^{(\ell)}] ,$\newline
where $B^{(k)}_{i} = (C^{(k)}L^{i})_{+}$.  For example, we have:

$$B^{(k)}_{1} = E_{kk}\partial + C^{(k,1)},\ B^{(k)}_{2} = E_{kk}\partial^{2} +
C^{(k,1)}\partial + 2E_{kk} U^{(1)} + C^{(k,2)}. \tag{3.8.4}$$

Denote by $C^{(k,\ell)}_{ij}$ and $U^{(k)}_{ij}$ the $(i,j)$-th
entries of the $n \times n$ matrices $C^{(k,\ell)}$ and $U^{(k)}$
respectively.  Then the $\partial^{-1}$ term of the second equation
(3.8.2) gives:

$$C^{(k,1)}_{ij} = 0 \ \text{if}\ i \neq k\ \text{and}\ j \neq k,\ \text{or}\ i
= j = k, \tag{3.8.5}$$

$$C^{(k,1)}_{kj} = -C^{(j,1)}_{kj}. \tag{3.8.6}$$
Hence the matrices $C^{(j,1)}$ are expressed in terms of the
functions

$$A_{ij} := C^{(j,1)}_{ij}\ \ (\text{note that}\ A_{ii} = 0).$$
The $\partial^{-2}$ term of the second equation  (3.8.2) allowes
one to express most of the $C^{(k,2)}_{ij}$ in terms of the $A_{ij}$:

$$C^{(k,2)}_{ij} = -A_{ik}A_{kj}\ \text{if}\ i \neq k\ \text{and}\ j
\neq k, \tag{3.8.7}$$

$$C^{(k,2)}_{k,k} = \sum^{n}_{p=1} A_{kp} A_{pk}. \tag{3.8.8}$$

Furthermore, the $\partial^{-1}$ term of the Lax equation
(3.8.3b)$_{1}$ gives:

$$\frac{\partial A_{ij}}{\partial x^{(k)}_{1}} = A_{ik} A_{kj}\
\text{for distinct}\ i,j,k, \tag{3.8.9}$$

$$C^{(j,2)}_{ij} = - \frac{\partial A_{ij}}{\partial x^{(j)}_{1}}\
\text{for} \ i \neq  j, \tag{3.8.10}$$

$$C^{(i,2)}_{ij} = \sum^{n}_{\Sb p=1 \\ p \neq i\endSb}
\frac{\partial A_{ij}}{\partial x^{(p)}_{1}} \ \text{for} \ i \neq j.
\tag{3.8.11}$$
The $\partial^{-2}$ term of that equation gives for $i \neq j$
(recall that $\partial = \frac{\partial}{\partial x^{(1)}_{1}} +
\ldots +  \frac{\partial}{\partial x^{(n)}_{1}}$.):

$$C^{(i,3)}_{ij} = -
\frac{\partial C^{(i,2)}_{ij}}{\partial x^{(j)}_{1}} + A_{ij}
C^{(i,2)}_{jj} - \sum^{n}_{p=1}
(A_{ip} \partial A_{pj} + C^{(i,2)}_{ip} A_{pj}) , \tag{3.8.12}$$

$$C^{(j,3)}_{ij} = - \sum^{n}_{\Sb p = 1 \\ p \neq i\endSb}
\frac{\partial C^{(j,2)}_{ij}}{\partial
x^{(p)}_{1}} - A_{ij} C^{(j,2)}_{ii} +
\sum^{n}_{p=1} A_{ip} C^{(j,2)}_{pj} .
\tag{3.8.13}$$

Substituting (3.8.7, 8 and 11) (resp. (3.8.7, 8 and 10)) in (3.8.12)
(resp. in (3.8.13)) we obtain for $i \neq j$:

$$C^{(i,3)}_{ij} = - (\partial - \frac{\partial}{\partial
x^{(i)}_{1}})^{2} A_{ij} - 2 \sum^{n}_{\Sb p = 1 \\ p \neq i\endSb}
A_{ip} A_{pi} A_{ij} , \tag{3.8.14}$$

$$C^{(j,3)}_{ij} = \frac{\partial^{2} A_{ij}}{\partial x^{(j)2}_{1}}
+ 2 \sum^{n}_{\Sb p = 1 \\ p \neq j\endSb} A_{ij} A_{jp} A_{pj} .
\tag{3.8.15}$$

Furthermore, the $\partial^{0}$ and $\partial^{-1}$ terms of the Lax
equation (3.8.3a)$_{1}$ give respectively for $i \neq j$:

$$U^{(1)}_{ij} = -\partial A_{ij}, \tag{3.8.16}$$

$$\frac{\partial U^{(1)}_{ii}}{\partial x^{(j)}_{1}} = - \partial
(A_{ij}A_{ji}). \tag{3.8.17}$$

Finally, the $\partial^{-1}$ term of the Lax equation (3.8.3b)$_{2}$
gives

$$\frac{\partial A_{ij}}{\partial x^{(j)}_{2}} = -2A_{ij}U^{(1)}_{jj}
- C^{(j,3)}_{ij}\ \text{for}\ i \neq j, \tag{3.8.18}$$

$$\frac{\partial A_{ij}}{\partial x^{(i)}_{2}} = \partial^{2}A_{ij} -
2\partial C^{(i,2)}_{ij} - C^{(i,3)}_{ij} + 2U^{(1)}_{ii}A_{ij}\
\text{for}\ i \neq j, \tag{3.8.19}$$

$$\frac{\partial A_{ij}}{\partial x^{(k)}_{2}} = A_{ik}
\frac{\partial A_{kj}}{\partial x^{(k)}_{1}} - A_{kj} \frac{\partial
A_{ik}}{\partial x^{(k)}_{1}} \ \text{for}\ i \neq k\ \text{and} \ j
\neq k. \tag{3.8.20}$$

\vskip 10pt
{\bf 3.9.}  Finally, we write down explicitly expressions for
$U^{(1)}$ and $C^{(i,1)}$ in terms of $\tau$-functions.  Recall that

$$\align
P &= I_{n} + \sum^{\infty}_{j=1} W^{(j)}(x)\partial^{-j}, \\
L &= P\partial P^{-1} = I_{n} \partial + \sum^{\infty}_{j=1}  U^{(j)}
\partial^{-j}, \\
C^{(i)} &= PE_{ii}P^{-1} = E_{ii} + \sum^{\infty}_{j=1}
C^{(i,j)}\partial^{-j}.\endalign$$
Using (3.3.18) we have:

$$U^{(1)} = -\partial W^{(1)}, \tag{3.9.1}$$

$$U^{(2)} = W^{(1)} \partial W^{(1)} - \partial W^{(2)},
\tag{3.9.2}$$

$$C^{(i,1)} = [W^{(1)},E_{ii}], \tag{3.9.3}$$

$$C^{(i,2)} = [W^{(2)},E_{ii}] + [E_{ii},W^{(1)}]W^{(1)}.
\tag{3.9.4}$$
Using (3.3.7) we obtain from (3.9.1) and (3.9.3) respectively:

$$U^{(1)}_{ij} = \cases -\varepsilon_{ji} \partial (\tau_{\alpha +
\alpha_{ij}}/\tau_{\alpha}) &\text{if}\ i
\neq j, \\
\partial (\frac{\partial \tau_{\alpha}}{\partial
x^{(i)}_{1}}/\tau_{\alpha}) &\text{if}\ i = j. \endcases
\tag{3.9.5}$$

$$A_{ij} \equiv C^{(j,1)}_{ij} = \varepsilon_{ji} \tau_{\alpha +
\alpha_{ij}}/\tau_{\alpha} . \tag{3.9.6}$$
(Recall that by (3.8.5 and 6) all the matrices $C^{(k,1)}$ can be
expressed via the functions $A_{ij}$.)  Using (3.3.7 and 8) and
(3.9.2 and 4) one also may write down the matrices $U^{(2)}$ and
$C^{(i,2)}$ in terms of $\tau$-functions, but they are somewhat more
complicated and we will not need them anyway.

\vskip 10pt

\subheading{\S 4.  The $n$-wave interaction equations, the generalized
Toda chain and the generalized
Davey-Stewartson equations as subsystems of the $n$-component KP}

\vskip 10pt

{\bf 4.0.} In this section we show that some well-known soliton
equations, as well as their natural generalizations, are the simplest
equations of the various formulations of the $n$-component KP
hierarchy.  To simplify notation, let

$$t_{i} = x^{(i)}_{2},\ x_{i} = x^{(i)}_{1}, \ \text{so that}\
\partial = \sum^{n}_{i=1} \frac{\partial}{\partial x_{i}}.
\tag{4.0.1}$$

{\bf 4.1.}  Let $n \geq 3$.  Then the $n$-component KP in the form
of Sato equation contains the system (3.7.1) of $n(n-1)(n-2)$
equations on $n^{2}-n$ functions $W_{ij}\ (i \neq j)$ in the
indeterminates $x_{i}$ (all other indeterminates being parameters):

$$\frac{\partial W_{ij}}{\partial x_{k}} = W_{ik} W_{kj}\ \text{for
distinct}\ i,j,k.\ \tag{4.1.1}$$
The $\tau$-function is given by the formula (3.3.7) for a fixed
$\alpha \in M$:

$$W_{ij} = \varepsilon_{ji}\ \tau_{\alpha + \alpha_{ij}} /
\tau_{\alpha} . \tag{4.1.2}$$
Substituting this in (4.1.1) gives the Hirota bilinear equation
(2.4.8):

$$D^{(k)}_{1} \tau_{\alpha} \cdot \tau_{\alpha + \alpha_{ij}} =
\varepsilon_{ik} \varepsilon_{kj} \varepsilon_{ij} \tau_{\alpha +
\alpha_{ik}} \tau_{\alpha + \alpha_{kj}} \tag{4.1.3}$$
Note that due to (3.9.3), $W_{ij} = A_{ij}$ if $i \neq j$, hence
(4.1.1) is satisfied by the $A_{ij}$ as well.

One usually adds to (4.1.1) the equations

$$\partial W_{ij} = 0 ,\ i \neq j. \tag{4.1.4}$$
We shall explain the group theoretical meaning of this constraint in
\S 6.

Let now $a = \text{diag} (a_{1},\ldots ,a_{n}),\ b = \text{diag}
(b_{1},\ldots ,b_{n})$ be arbitrary diagonal matrices over ${\Bbb
C}$.  We reduce the system (4.1.1) to the plane [D]:

$$x_{k} = a_{k}x + b_{k}t. \tag{4.1.5}$$
A direct calculation shows that (4.1.1) reduces then to the following
equation on the matrix $W = (W_{ij})$ (note that its diagonal entries
don't occur):

$$[a , \frac{\partial W}{\partial t}] - [b , \frac{\partial
W}{\partial x}] = [[a,W],[b,W]] + b\partial Wa - a\partial Wb.
\tag{4.1.6}$$
Hence, imposing the constraint (4.1.4), we obtain the famous $1+1$
$n$-wave system (cf. [D], [NMPZ]):

$$[a,\frac{\partial W}{\partial t}] - [b, \frac{\partial W}{\partial
x}] = [[a,W],[b,W]]. \tag{4.1.7}$$

Let now

$$x_{k} = a_{k}x + b_{k}t - y . \tag{4.1.8}$$
Then equation (4.1.6) gives

$$[a,\frac{\partial W}{\partial t}] - [b , \frac{\partial W}{\partial
x}] - a \frac{\partial W}{\partial y} b + b \frac{\partial
W}{\partial y} a = [[a,W],[b,W]]. \tag{4.1.9}$$
If we let

$$Q_{ij} = -(a_{i}-a_{j})W_{ij}. $$
equation (4.1.9) turns into the following system, which is called in
[AC, (5.4.30a,c)] the $2+1$ $n$-wave interaction equations $(i \neq j)$:

$$\frac{\partial Q_{ij}}{\partial t} = a_{ij} \frac{\partial
Q_{ij}}{\partial x} + b_{ij} \frac{\partial Q_{ij}}{\partial y} +
\sum_{k} (a_{ik} - a_{kj}) Q_{ik} Q_{kj} , \tag{4.1.10}$$
where

$$a_{ij} = (b_{i}-b_{j})/(a_{i}-a_{j}),\ b_{ij} = b_{i} - a_{i}a_{ij}.
\tag{4.1.11}$$

On the other hand, letting (we assume that $a_{1} > \ldots > a_{n}$):

$$w_{ij} = W_{ij}/(a_{i}-a_{j})^{1/2},\tag{4.1.12}$$
the equation (4.1.6) gives for $i \neq j$:

$$\frac{\partial w_{ij}}{\partial t} - a_{ij} \frac{\partial
w_{ij}}{\partial x} - b_{ij} \frac{\partial w_{ij}}{\partial y} =
\sum_{k} \varepsilon_{ijk} w_{ik} w_{kj} , \tag{4.1.13}$$
where

$$\varepsilon_{ijk} = \frac{a_{i}b_{k} + a_{k}b_{j} + a_{j}b_{i} -
a_{k}b_{i}-a_{j}b_{k} -
a_{i}b_{j}}{((a_{i}-a_{k})(a_{k}-a_{j})(a_{i}-a_{j}))^{1/2}} .
\tag{4.1.14}$$
Imposing the constraint $\overline{w}_{ij} = -w_{ji}$, we obtain from
(4.1.13) the following Hamiltonian system (considered in [NMPZ, pp 175,
242] for $n = 3$ and called there the $2+1$ $3$-wave system)  $(i < j)$:

$$\frac{\partial w_{ij}}{\partial t} - a_{ij} \frac{\partial
w_{ij}}{\partial x} - b_{ij} \frac{\partial w_{ij}}{\partial y} =
\frac{\partial H}{\partial \overline{w}_{ij}} , \tag{4.1.15}$$
where
$$H = \sum \Sb i,k,j \\ i < k < j\endSb \varepsilon_{ijk}
(w_{ik}w_{kj}\overline{w}_{ij} + \overline{w}_{ik} \overline{w}_{kj}
w_{ij}). \tag{4.1.16}$$

Finally, let $n = 3$ and let $u_{1} = iw_{13},\ u_{2} =
i\overline{w}_{13},$  $u_{3} = iw_{12},$ $a_{1} = -a_{23},$ $b_{1} =
-b_{23},$ $a_{2} = -\overline{a}_{13},$ $b_{2} = -\overline{b}_{13},$
$a_{3} = -a_{12},$ $b_{3} = -b_{13}$.  Then, after imposing the
constraint $\varepsilon_{132} = 1$, equations
(4.1.15) turn into the well-known $2 + 1$\ $3$-wave interaction equations
(see [AC, (5.4.27)]):

$$\frac{\partial u_{j}}{\partial t} + a_{j} \frac{\partial
u_{j}}{\partial x} + b_{j} \frac{\partial u_{j}}{\partial y} =
i\overline{u}_{k} \overline{u}_{\ell} , \tag{4.1.17}$$
where $(j,k,\ell )$ is an arbitrary cyclic permutation of $1,2,3$.

\vskip 10pt
{\bf 4.2.}  Let $n \geq 2$.  Then the $n$-component KP in the form
of Sato equations contains the following subsystem of the system of
equations (3.7.1) for arbitrary $\alpha \in M$ on the functions
$W_{ij}(\alpha)$ in the indeterminates $x_{i}$ (all other
indeterminates being parameters):

$$\frac{\partial W_{ii}(\alpha )}{\partial x_{j}} = W_{ij} (\alpha )
W_{ji} (\alpha )\ \text{if}\ i \neq j . \tag{4.2.1}$$
The $\tau$-function is given by (3.3.7) $(\alpha \in M)$:

$$W_{ij}(\alpha ) = \cases \varepsilon_{ji} \tau_{\alpha +
\alpha_{ij}}/\tau_{\alpha } &\text{if}\ i \neq j \\
- \frac{\partial}{\partial x_{i}} \log \tau_{\alpha} &\text{if}\ i =
j . \endcases \tag{4.2.2}$$
Substituting this in (4.2.1) gives the Hirota bilinear equations
(2.4.4):

$$D_{i}D_{j} \tau_{\alpha} \cdot \tau_{\alpha } = 2\tau_{\alpha +
\alpha_{ij}} \tau_{\alpha - \alpha_{ij}} . \tag{4.2.3}$$
In order to rewrite (4.2.1) in a more familiar form, let for $i \neq
j$:

$$U_{ij}(\alpha ) = \log \varepsilon_{ji} W_{ij}(\alpha ) = \log
(\tau_{\alpha + \alpha_{ij}} /\tau_{\alpha}). \tag{4.2.4}$$
Note that $\log (\tau_{\alpha + \alpha_{ij}} /\tau_{\alpha }) = -\log
(\tau_{(\alpha + \alpha_{ij})-\alpha_{ij}} / \tau_{\alpha +
\alpha_{ij}}$.  Hence from (4.2.2) we obtain

$$U_{ij}(\alpha ) = -U_{ji}(\alpha + \alpha_{ij})\ \text{if}\ i \neq
j . \tag{4.2.5}$$
Furthermore, we have:

$$\align
\frac{\partial^{2}}{\partial x_{i} \partial x_{j}} U_{ij}(\alpha ) &=
\frac{\partial^{2}}{\partial x_{i} \partial x_{j}} \log \tau_{\alpha
+ \alpha_{ij}} - \frac{\partial^{2}}{\partial x_{i}\partial x_{j}}
\log \tau_{\alpha} \\
&= \frac{\partial W_{ii}(\alpha )}{\partial x_{j}} - \frac{\partial
W_{ii} (\alpha + \alpha_{ij})}{\partial x_{j}}
= W_{ij}(\alpha ) W_{ji} (\alpha ) - W_{ij} (\alpha +
\alpha_{ij})W_{ji}(\alpha + \alpha_{ij}) \\
&= - \frac{\tau_{\alpha + \alpha_{ij}}}{\tau_{\alpha}}
\frac{\tau_{\alpha - \alpha_{ij}}}{\tau_{\alpha}} + \frac{\tau_{\alpha
+ 2\alpha_{ij}}}{\tau_{\alpha + \alpha_{ij}}}
\frac{\tau_{\alpha}}{\tau_{\alpha + \alpha_{ij}}}
= e^{U_{ij}(\alpha + \alpha_{ij})-U_{ij}(\alpha)} - e^{U_{ij}(\alpha
)-U_{ij}(\alpha - \alpha_{ij})} .
\endalign$$
Thus the functions $U_{ij}(\alpha )\  (i \neq j)$ satisfy the
following generalized Toda chain (with constraint (4.2.5)):

$$\frac{\partial^{2}U_{ij}(\alpha )}{\partial x_{i}\partial x_{j}} =
e^{U_{ij}(\alpha + \alpha_{ij}) - U_{ij}(\alpha )} -
e^{U_{ij}(\alpha ) - U_{ij} (\alpha - \alpha_{ij})} . \tag{4.2.6}$$
Note also that (4.1.1) for distinct $i,j$ and $k$ becomes:

$$\frac{\partial U_{ij}(\alpha )}{\partial x_{k}} = \varepsilon_{ik}
\varepsilon_{kj} \varepsilon_{ji} e^{U_{ij}(\alpha ) + U_{ik}(\alpha
) + U_{kj}(\alpha )} . \tag{4.2.7}$$

One should be careful about the boundary conditions.  Let $S = \
\text{supp}\ \tau$; recall that by Proposition 2.4, $S$ is a convex
polyhedron with vertices in $M$ and edges parallel to roots.  It
follows that (4.2.6) should be understood as follows:

(i) if $\alpha \notin S$, then $U_{ij}(\alpha ) = 0$ and
(4.2.6) is trivial,

(ii) if $\alpha \in S$, but $\alpha + \alpha_{ij} \notin
S$, then (4.2.6) is trivial,

(iii) if $\alpha \in S$, but $\alpha - \alpha_{ij}
\notin S$, then the second term on the right-hand side of (4.2.6) is
removed,

(iv) if $\alpha \in S,\ \alpha + \alpha_{ij} \in S$, but
$\alpha + 2 \alpha_{ij} \notin S$, then the first term on the
right-hand side of (4.2.6) is removed.

Let now $n = 2$, and let $u_{n} = U_{12}(n\alpha_{12})$.  Then we get
the usual Toda chain:

$$\frac{\partial^{2}u_{n}}{\partial x_{1}\partial x_{2}} =
e^{u_{n+1}-u_{n}} - e^{u_{n}-u_{n-1}},\ n \in {\Bbb Z}. \tag{4.2.8}$$
It is a part of the Toda lattice hierarchy discussed in [UT].

\vskip 10pt
{\bf 4.3.} Let $n \geq 2$.  Then the $n$-component KP in the form of
Sato equations contains the system of equations (3.7.4), (3.7.5),
(3.7.3) and (3.7.1) for $j \neq k$ on $n^{2}$ functions $W_{ij}$ in
the indeterminates $x_{k}$ and $t_{k}$ $(k = 1,\ldots ,n)$ (all other
indeterminates being parameters):

$$\frac{\partial W_{ij}}{\partial t_{j}} = -
\frac{\partial^{2}W_{ij}}{\partial x^{2}_{j}} + 2 \frac{\partial
W_{jj}}{\partial x_{j}} W_{ij} \ \text{if}\ i \neq j, \tag{4.3.1}$$

$$\frac{\partial W_{ij}}{\partial t_{i}} =
\frac{\partial^{2}W_{ij}}{\partial x^{2}_{i}} - 2 \frac{\partial
W_{ii}}{\partial x_{i}} W_{ij}\ \text{if} \ i \neq j, \tag{4.3.2}$$

$$\frac{\partial W_{ij}}{\partial t_{k}} = W_{ik} \frac{\partial
W_{kj}}{\partial x_{k}} - \frac{\partial W_{ik}}{\partial x_{k}}
W_{kj}\ \text{if}\ i \neq k\ \text{and}\ j \neq k, \tag{4.3.3}$$

$$\frac{\partial W_{ij}}{\partial x_{k}} = W_{ik} W_{kj}\ \text{if}\
i \neq k\ \text{and}\ j \neq k. \tag{4.3.4}$$
This is a system of $n^{3}-n$ evolution equations (4.3.1--3) and
$n(n-1)^{2}$ constraints (4.3.4) which we call the generalized
Davey-Stewartson system.

Note that the $\tau$-functions of this system are given by (3.3.7),
where we may take $\alpha = 0$.  The corresponding to
(4.3.1)--(4.3.4) Hirota bilinear equations are (2.4.6); (2.4.5);
(2.4.7) if $i = j$ and (2.4.9) if $i \neq j$; (2.4.4) if $i = j$ and
(2.4.8) if $i \neq j$, respectively.

Now, note that letting

$$\varphi_{ij} = \frac{1}{2} \left( \frac{\partial W_{ii}}{\partial
x_{i}} + \frac{\partial W_{jj}}{\partial x_{j}} + \frac{\partial
W_{ii}}{\partial x_{j}} + \frac{\partial W_{jj}}{\partial
x_{i}}\right) (= \varphi_{ji}),$$
and subtracting (4.3.2) from (4.3.1) we obtain using (4.3.4):

$$\frac{\partial W_{ij}}{\partial t_{j}} - \frac{\partial
W_{ij}}{\partial t_{i}} = -  \left( \frac{\partial^{2}}{\partial
x^{2}_{i}} + \frac{\partial^{2}}{\partial x^{2}_{j}}\right) W_{ij} +
2W_{ij} (\varphi_{ij} - W_{ij}W_{ji}). \tag{4.3.5}$$
Also, from (4.3.4) we obtain

$$\frac{\partial^{2}\varphi_{ij}}{\partial x_{i}\partial x_{j}} =
\frac{1}{2} \left( \frac{\partial}{\partial x_{i}} +
\frac{\partial}{\partial x_{j}}\right)^{2} (W_{ij}W_{ji}).
\tag{4.3.6}$$

Let now $n = 2$; to simplify notation, let

$$q = W_{12},\ r = W_{21},\ \varphi = \varphi_{12} = \varphi_{21}.$$

Then, making the change of indeterminates

$$s = -2i (t_{1} + t_{2}) ,\ t = -2i(t_{1}-t_{2}),\ x= x_{1} +
x_{2},\ y = x_{1} - x_{2}, \tag{4.3.7}$$
equations (4.3.5 and 6) turn into the decoupled Davey-Stewartson
system:

$$\cases i \frac{\partial q}{\partial t}  = - \frac{1}{2}
(\frac{\partial^{2}q}{\partial x^{2}} + \frac{\partial^{2}q}{\partial
y^{2}}) +  q(\varphi - qr) \\
i \frac{\partial r}{\partial t} = \frac{1}{2}
(\frac{\partial^{2}r}{\partial x^{2}} + \frac{\partial^{2}r}{\partial
y^{2}}) - r(\varphi - qr) \\
\frac{\partial^{2}\varphi}{\partial x^{2}} -
\frac{\partial^{2}\varphi}{\partial y^{2}} = 2
\frac{\partial^{2}(qr)}{\partial x^{2}} . \endcases \tag{4.3.8}$$
Due to (3.3.7), the corresponding $\tau$-functions are given by the
following formulas, where we let $\tau_{n} = \tau_{n\alpha_{12}}$:

$$q = - \tau_{1}/\tau_{0},\ r = \tau_{-1}/\tau_{0},\ \varphi =
-\frac{\partial^{2}}{\partial x^{2}} \log \tau_{0}, \tag{4.3.9}$$
the Hirota bilinear equations being (cf. [HH]):

$$\aligned
&(iD_{t} + \frac{1}{2} D^{2}_{x} + \frac{1}{2} D^{2}_{y})\tau_{1}
\cdot \tau_{0} = 0 \\
&(-iD_{t} + \frac{1}{2} D^{2}_{x} + \frac{1}{2} D^{2}_{y})\tau_{-1}
\cdot \tau_{0} = 0 \\
&(D^{2}_{x} - D^{2}_{y})\tau_{0} \cdot \tau_{0} = 2\tau_{1}\tau_{-1} .
\endaligned \tag{4.3.10}$$

Finally, imposing the constraint

$$r = \kappa\overline{q},\ \text{where}\ \kappa = \pm 1 , \tag{4.3.11}$$
we obtain the classical Davey-Stewartson system

$$\cases i \frac{\partial q}{\partial t} + \frac{1}{2}
(\frac{\partial^{2}q}{\partial x^{2}} + \frac{\partial^{2}q}{\partial
y^{2}}) = (\varphi - \kappa |q|^{2})q & \ \\
\frac{\partial^{2}\varphi}{\partial x^{2}} -
\frac{\partial^{2}\varphi}{\partial y^{2}} = 2\kappa
\frac{\partial^{2}|q|^{2}}{\partial x^{2}}. \endcases \tag{4.3.12}$$

\vskip 10pt
\noindent
REMARK 4.3.  It is interesting to compare the above results with that
obtained via the Lax equations.  To simplify notation, let $U_{i} =
U^{(1)}_{ii}$.  Substituting (3.8.15) (resp. (3.8.14)) in (3.8.18)
(resp. (3.8.19)), we obtain for $i \neq j$:

$$\frac{\partial A_{ij}}{\partial t_{j}} = -
\frac{\partial^{2}A_{ij}}{\partial x^{2}_{j}} - 2 A_{ij}U_{j} - 2
\sum_{k \neq j} A_{ij} A_{jk} A_{kj} \tag{4.3.13}$$

$$\frac{\partial A_{ij}}{\partial t_{i}} =
\frac{\partial^{2}A_{ij}}{\partial x^{2}_{i}} + 2 A_{ij} U_{i} + 2
\sum_{k \neq i} A_{ij} A_{ik} A_{ki} \tag{4.3.14}$$
These equation together with (3.8.9, 17 and 20) give a slightly
different version of the generalized DS system (recall that $A_{ij} =
W_{ij}$ if $i \neq j$ and $U_{i} = - \partial W_{ii}$).  For $n = 2$
we get again the classical DS system after the change of
indeterminates (4.3.7) if we let $\varphi = - \frac{1}{2} (U_{1} + U_{2})$.

\vskip 10pt
{\bf 4.4.}  Finally, we explain what happens in the well-known case
$n = 1$.  In this case $C^{(1)} = 1$ and auxiliary conditions (3.8.2)
are trivial.  Lax equation (3.8.3b) is trivial as well, and Lax
equation (3.8.3a) becomes

$$\frac{\partial L}{\partial x_{i}} = [B_{i},L],\ i = 1,2,\ldots ,
\tag{4.4.1}$$
where $L = \partial + \sum^{\infty}_{j=1} u_{j}(x) \partial^{-j},\
\partial = \frac{\partial}{\partial x_{1}}$ and $B_{i} =
(L^{i})_{+}$.  Thus, the KP hierarchy is a system of partial
differential equations (4.4.1) on unknown functions
$u_{1},u_{2},\ldots $ in indeterminates $x_{1},x_{2},\ldots $.  By
Lemma 3.6, (4.4.1) is equivalent to the following system of
Zakharov-Shabat equations:

\vskip 10pt
\noindent
(4.4.2)$_{k,\ell}$ \ \ \ \ \ \ \ \ \ \ $\frac{\partial B_{\ell}}{\partial
x_{k}} - \frac{\partial B_{k}}{\partial x_{\ell}} = [B_{k},B_{\ell}].$
\vskip 5pt
\noindent
By (3.8.4) we have:

$$B_{1} = \partial ,\ B_{2} = \partial^{2} + 2u_{1}. \tag{4.4.3}$$
Furthermore, we have:

$$B_{3} = \partial^{3} + 3u_{1}\partial + 3u_{2} + 3 \frac{\partial
u_{1}}{\partial x_{1}} \tag{4.4.4}$$
Thus we see that equations (4.4.2)$_{k,1}$ are all trivial, the first
non-trivial equation of (4.4.2) being

$$\frac{\partial B_{2}}{\partial x_{3}} - \frac{\partial
B_{3}}{\partial x_{2}} = [B_{2},B_{3}].$$
Substituting in it (4.4.3 and 4), the coefficients of $\partial^{0}$
and $\partial^{1}$ give respectively:

$$2 \frac{\partial u_{1}}{\partial x_{3}} - 2
\frac{\partial^{2}u_{1}}{\partial x_{1}\partial x_{2}} - 6u_{1}
\frac{\partial u_{1}}{\partial x_{1}} = 3 \frac{\partial
u_{2}}{\partial x_{2}} - 3 \frac{\partial^{2}u_{2}}{\partial
x^{2}_{1}} \tag{4.4.5}$$
$$6 \frac{\partial u_{2}}{\partial x_{1}} = 3 \frac{\partial
u_{1}}{\partial x_{2}} - \frac{\partial^{2}u_{1}}{\partial x^{2}_{1}}
. \tag{4.4.6}$$
Differentiating (4.4.5) by $x_{1}$ and substituting $\frac{\partial
u_{2}}{\partial x_{2}}$ from (4.4.6) gives a PDE on $u = 2u_{1}$,
where we let $x_{1} = x,\ x_{2} = y, x_{3} = t$:

$$\frac{3}{4} \frac{\partial^{2}u}{\partial y^{2}} =
\frac{\partial}{\partial x} \left(\frac{\partial u}{\partial t} -
\frac{3}{2} u \frac{\partial u}{\partial x} - \frac{1}{4}
\frac{\partial^{3}u}{\partial x^{3}} \right) . \tag{4.4.7}$$
This is the classical KP equation.  Due to (3.9.5), the connection
between $u$ and the $\tau$-function is given by the famous formula

$$u = 2 \frac{\partial^{2}}{\partial x^{2}} \log \tau . \tag{4.4.8}$$
Substituting $u$ in (4.4.7) gives the Hirota bilinear equation
(2.4.3).
\vskip 10pt

\subheading{\S 5. Soliton and dromion solutions}
\vskip 10pt
{\bf 5.1.}  We turn now to the construction of solutions of the
$n$-component KP hierarchy.  As in [DJKM3] we make use of the vertex
operators (2.1.14).  When transported via the $n$-component
boson-fermion correspondence $\sigma$ from $F$ to $B = {\Bbb C}[x]
\otimes {\Bbb C}[L]$, they take the following form:

$$\psi^{\pm (i)}(z) = Q^{\pm 1}_{i} z^{\pm \alpha^{(i)}_{0}} (\exp
\pm \sum^{\infty}_{k=1} z^{k}x^{(i)}_{k}) (\exp \mp
\sum^{\infty}_{k=1} \frac{z^{-k}}{k} \frac{\partial}{\partial
x^{(i)}_{k}}). \tag{5.1.1}$$
Note that for $z,w \in {\Bbb C}^{\times}$ such that $|w| < |z|$ we
have $(\lambda ,\mu = +\ \text{or}\ -)$:

$$\aligned
&\psi^{\lambda (i)}(z) \psi^{\mu (j)}(w) = (z-w)^{\delta_{ij}\lambda
\mu} Q^{\lambda 1}_{i}  Q^{\mu 1}_{j}
z^{\alpha^{(i)}_{0}} w^{\alpha ^{(j)}_{0}} \\
& \times \exp \sum^{\infty}_{k=1} (\lambda z^{k}x^{(i)}_{k} + \mu
w^{k}x^{(j)}_{k}) \exp -\sum^{\infty}_{k=1} (\lambda \frac{z^{-k}}{k}
\frac{\partial}{\partial x^{(i)}_{k}} + \mu \frac{w^{-k}}{k}
\frac{\partial}{\partial x^{(j)}_{k}}).
\endaligned \tag{5.1.2}
$$
We let for $0 < |w| < |z|$:

$$\aligned
&\Gamma_{ij} (z,w) \overset{\text{def}}\to{=} \psi^{+(i)}(z)
\psi^{-(j)}(w) = (z-w)^{-\delta_{ij}}Q_{i}Q^{-1}_{j}
z^{\alpha^{(i)}_{0}} w^{-\alpha^{(j)}_{0}} \exp \sum^{\infty}_{k=1}
(z^{k}x^{(i)}_{k} - w^{k}x^{(j)}_{k})\\
& \times  \exp - \sum^{\infty}_{k=1}
(\frac{z^{-k}}{k} \frac{\partial}{\partial x^{(i)}_{k}} -
\frac{w^{-k}}{k} \frac{\partial}{\partial x^{(j)}_{k}}) .
\endaligned\tag{5.1.3}$$

Using (5.1.2), we obtain for $|z_{1}| > |z_{2}| > \ldots > |z_{2N-1}|
> |z_{2N}| > 0$:

$$\aligned
&\Gamma_{i_{1}i_{2}} (z_{1},z_{2})\ldots
\Gamma_{i_{2N-1}i_{2N}}(z_{2N-1},z_{2N}) = \prod_{1 \leq k < \ell
\leq 2N} (z_{k}-z_{\ell})^{(-1)^{k+\ell}\delta_{i_{k}i_{\ell}} } \\
&\times Q_{i_{1}} Q^{-1}_{i_{2}} \ldots Q_{i_{2N-1}}Q^{-1}_{i_{2N}}
\prod^{2N}_{m=1} z^{(-1)^{m-1}\alpha^{(i_{m})}_{0}}_{m} \exp
(-\sum^{2N}_{m=1} \sum^{\infty}_{k=1} (-1)^{m} z^{k}_{m} x^{(m)}_{k}) \\
& \times \exp ( \sum^{2N}_{m=1} \sum^{\infty}_{k=1} (-1)^{m}
\frac{z^{-k}_{m}}{k} \frac{\partial}{\partial x^{(m)}_{k}}).
\endaligned \tag{5.1.4}
$$
We may analytically extend the right-hand side of (5.1.4) to the
domain $\{ z_{i} \neq 0,\ z_{i} \neq z_{j}\ \text{if}\ i \neq j,\ i,j
= 1,\ldots ,2N\}.$  Then we deduce from (5.1.4) for $N = 2$ that in
this domain we have:

$$\Gamma_{i,i_{2}} (z_{1},z_{2}) \Gamma_{i_{3},i_{4}} (z_{3},z_{4}) =
\Gamma_{i_{3}i_{4}} (z_{3},z_{4}) \Gamma_{i,i_{2}}(z_{1},z_{2}),
\tag{5.1.5}$$

$$\Gamma_{ij}(z_{1},z_{2})^{2} \equiv \lim_{\Sb z_{3} @>>> z_{1} \\
z_{4} @>>> z_{2}\endSb} \Gamma_{ij} (z_{1},z_{2}) \Gamma_{ij}(z_{3},z_{4})
= 0. \tag{5.1.6}$$

\noindent
REMARK 5.1.  Let $A = (a_{ij})$ be a $n \times n$ matrix over ${\Bbb
C}$ and let $z_{i},w_{i}\ (i = 1,\ldots ,n)$ be non-zero complex
numbers such that $z_{i} \neq w_{j}$.  Due to (1.2.3) the sum

$$\Gamma_{A}(z,w) = \sum^{n}_{i,j=1} a_{ij} \Gamma_{ij} (z_{i},w_{j})
\tag{5.1.7}$$
lies in a completion of $r(g\ell_{\infty})$.

By (5.1.5--6) we obtain:

$$\exp \Gamma_{A}(z,w) = \prod^{n}_{i,j=1} (1 + a_{ij}
\Gamma_{ij}(z_{i},w_{j})). \tag{5.1.8}$$

\proclaim{Lemma 5.1} (a) If $\tau$ is a solution of the
$n$-component KP hierarchy (2.3.7) of Hirota bilinear equations,
then $(\exp \Gamma_{A}(z,w))\tau$ is a solution as well for any
complex $n \times n$ matrix $A$ and any $z = (z_{1},\ldots z_{n})$,
$w = (w_{1},\ldots ,w_{n}) \in {\Bbb C}^{\times n}$ such that $z_{i} \neq
w_{j}$.

(b) For any collection of complex $n \times n$-matrices
$A_{1},\ldots ,A_{N}$ and any collection $z^{(1)}, \ldots,$
$z^{(N)}, w^{(1)}, \ldots, w^{(N)} \in {\Bbb C}^{\times n}$ with all
coordinates distinct, the function

$$\exp \Gamma_{A_{1}} (z^{(1)},w^{(1)}) \ldots \exp \Gamma_{A_{N}}
(z^{(N)},w^{(N)}) \cdot 1 \tag{5.1.9}$$
is a solution of the $n$-component KP hierarchy (2.3.7).
\endproclaim

\demo{Proof} (a) follows from Proposition 1.3 and Remark 5.1.  (b)
follows from (a) since the function $1 = \sigma|0\rangle$ satisfies
(1.3.1).\ \ \ \ \ \ $\square$
\enddemo

We call (5.1.9) the {\it $N$-solitary $\tau$-function} (of
the $n$-component KP hierarchy).  Note that the support of (5.1.9) can
be any polyhedron permitted by Proposition 2.4, which is in agreement
with the general belief that multisolitary solutions are dense in the
totality of all solutions.

In order to write down (5.1.9) in a more explicit form, introduce the
lexicographic ordering on the set $S$ of all triples $s = (p,i,j)$,
where $p \in \{1,\ldots ,N\},\ i,j \in \{1,\ldots ,n\}$ (i.e. $s_{1} <
s_{2}$ if $p_{1} < p_{2}$, or $p_{1} = p_{2}$ and $i_{1} < i_{2}$ or
$p_{1} = p_{2},\ i_{1} = i_{2}$ and $j_{1} < j_{2}$).  Given $N$ $n
\times n$ complex matrices $A_{p} = (a^{(p)}_{ij})$, we let $a_{s} =
a^{(p)}_{ij}$ for $s = (p,i,j) \in S$; given in addition two sets of
non-zero complex numbers $z_{s}$ and $w_{s}$, all distinct,
parametrized by $s \in S$, introduce the following constants

$$\aligned
&c(s_{1},\ldots ,s_{r}) = \prod^{r}_{k=1}
a_{s_{k}} \prod^{r}_{\ell = k+1} \varepsilon_{i_{k}i_{\ell}}
\varepsilon_{i_{k}j_{\ell}}
\varepsilon_{j_{k}i_{\ell}} \varepsilon_{j_{k}j_{\ell}}  \\
&\times \prod_{1 \leq k < \ell \leq r} \frac{
(z_{s_{k}}-z_{s_{\ell}})^{\delta_{i_{k}i_{\ell}}}
(w_{s_{k}}-w_{s_{\ell}})^{\delta_{j_{k}j_{\ell}}} }
{ (z_{s_{k}}-w_{s_{\ell}})^{\delta_{i_{k}j_{\ell}}}
(w_{s_{k}}-z_{s_{\ell}})^{\delta_{j_{k}i_{\ell}}} } .
\endaligned \tag{5.1.10}
$$
Then the $N$-solitary solution (5.1.9) can be written as follows

$$\aligned
&1 + \sum^{Nn^{2}}_{r=1} \sum_{(1,1,1) \leq s_{1} < \ldots < s_{r}
\leq (N,n,n)} c(s_{1},\ldots ,s_{r}) \\
& \times (\exp \sum^{r}_{k=1}
\sum^{\infty}_{m=1} (z^{m}_{s_{k}} x^{(i_{k})}_{m} - w^{m}_{s_{k}}
x^{(j_{k})}_{m})) e^{\sum^{r}_{k=1} \alpha_{i_{k}j_{k}}} .
\endaligned\tag{5.1.11}$$

\vskip 10pt
{\bf 5.2.}  Let $n = 1$.  Then the index set $S$ is naturally
identified with the set $\{1,\ldots ,N\}$, the two sets of complex
numbers we denote by $z_{2j-1}$ and $z_{2j}$, $j = 1,\ldots ,N$, and
we let $A_{p} = (z_{2p-1}-z_{2p})^{-1}a_{p}$ , where $a_{p}$ are some
constants.
Then (5.1.11) becomes the well-known formula (see [DJKM3]) for the
$\tau$-function of the $N$-soliton solution:

$$\aligned
\tau^{(N)} &= 1 + \sum^{N}_{r=1} \sum_{1 \leq j_{1} < \ldots < j_{r}
\leq N} \prod^{r}_{k=1} a_{j_{k}} \prod_{1 \leq k < \ell \leq 2r}
(z_{j_{k}}-z_{j_{\ell}})^{(-1)^{k+\ell}} \\
&\times \exp \sum^{r}_{k=1} \sum^{\infty}_{m=1} (z^{m}_{j_{2k-1}} -
z^{m}_{j_{2k}})x_{m} .
\endaligned \tag{5.2.1}
$$
Letting $x_{1} = x,\ x_{2} = y,\ x_{3} = t$ and all other
indeterminates constants $x_{4} = c_{4},\ldots$, we obtain, due to
(4.4.8), the soliton solution of the classical KP equation (4.4.7):

$$u(t,x,y) = 2 \frac{\partial^{2}}{\partial x^{2}} \log
\tau^{(N)}(x,y,t,c_{4},c_{5},\ldots ). \tag{5.2.2}$$

In particular, the $\tau$-function of the $1$-soliton solution is

$$\tau^{(1)}(x,y,t) = 1 + \frac{a}{z_{1}-z_{2}} \exp ((z_{1}-z_{2})x +
(z^{2}_{1} - z^{2}_{2})y + (z^{3}_{1} - z^{3}_{2})t + \
\text{const.}) \tag{5.2.3}$$
and we get the corresponding $1$-soliton solution of the classical
KP equation (4.4.7):

$$u(x,y,t) = \frac{(z_{1}-z_{2})^{2}}{2} \cosh^{-2} (\frac{1}{2}
((z_{1}-z_{2})x + (z^{2}_{1} - z^{2}_{2})y + (z^{3}_{1} -
z^{3}_{2})t) + \ \text{const.}). \tag{5.2.4}$$

\vskip 10pt
{\bf 5.3.}  Let $n = 2$.  Then any $\tau \in {\Bbb C}[x] \otimes
{\Bbb C}[M]$ can be written in the form

$$\tau = \sum_{\ell \in {\Bbb Z}} \tau_{\ell} e^{\ell \alpha_{12}},\
\text{where}\ \tau_{\ell} \equiv \tau_{\ell \alpha_{12}} .$$
For a $N$-solitary solution $\tau^{(N)}$ given by (5.1.11) we then
have

$$
\tau^{(N)}_{\ell} = \delta_{\ell ,0} + \sum^{4N}_{r=1}
\sum_{(s_{1},\ldots ,s_{r})} c(s_{1},\ldots ,s_{r})
 \exp \sum^{r}_{k=1} \sum^{\infty}_{m=1}
(z^{m}_{s_{k}}x^{(i_{k})}_{m} - w^{m}_{s_{k}} x^{(j_{k})}_{m}) ,
\tag{5.3.1}$$
where $(s_{1},\ldots ,s_{r})$ run over the subset (5.3.2)$_{2}$ of
$S^{r}$, where

\vskip 10pt
\noindent
(5.3.2)$_{n}$\ $\cases (1,1,1) \leq s_{1} < s_{2} \ldots < s_{r} \leq
(N,n,n) \\
\# \{ (i_{k},j_{k})|i_{k} > j_{k}\} - \# \{ (i_{k},j_{k})|i_{k} <
j_{k}\} = \ell . \endcases $
\vskip 5pt
\noindent
Letting (cf. (4.3.9)):
$$\aligned
q &= -
\frac{\tau_{1} (x,y,t,c,c^{(1)}_{3,\ldots})}
{\tau_{0} (x,y,t,c,c^{(1)}_{3},\ldots) }
,\ r = \frac{\tau_{-1}(x,y,t,c,c^{(1)}_{3},\ldots
)}{\tau_{0}(x,y,t,c,c^{(1)}_{3},\ldots )} , \\
\varphi &= - \frac{\partial^2}{\partial x^2} (\log
\tau_{0}(x,y,t,c,c^{(1)}_{3},\ldots )),
\endaligned \tag{5.3.3}$$
where $x = x^{(1)}_{1} + x^{(1)}_{2}, \ y = x^{(1)}_{1} -
x^{(1)}_{2},\ t = -2i(x^{(1)}_{2} - x^{(2)}_{2}),\ c =
-2i(x^{(1)}_{2} + x^{(2)}_{2})$ and all other indeterminates
$x^{(j)}_{k}$ are arbitrary constants $c^{(j)}_{k}$, we obtain a
$N$-solitary solution of the decoupled Davey-Stewartson system
(4.3.8).

We turn now to the classical Davey-Stewartson system (4.3.12) for
$\kappa = -1$.  The constraint (4.3.11) gives

$$\tau_{1}/\tau_{0} = \overline{\tau_{-1}/\tau_{0}}.$$
One way of satisfying this constraint is to let

$$\aligned
a^{(p)}_{ij} &= (-1)^{i+j}\overline{a}^{(p)}_{ji},\ z_{(p,i,j)} =
-\overline{w}_{(p,j,i)}, \\
c_{2} &= 0,\ c^{(j)}_{k} \in i^{k+1}{\Bbb R}.
\endaligned \tag{5.3.4}$$

We shall concentrate now on the case $N = 1$.  It will be convenient
to use the following notation:

$$\align
&x_{1} = x^{(1)}_{1},\ x_{2} = x^{(2)}_{1} , \\
&z_{ij} = z_{(1,i,j)} ,\ a_{i} = a_{(1,i,i)} \in {\Bbb R}_{+}\ (1
\leq i,j \leq 2),\ a_{3} = a_{(1,1,2)} \in {\Bbb C}, \\
&C(z,w) =
\frac{z-w}{z+\overline{w}},\ D(z,w) =
\frac{|a_{3}|^{2}}{(z+\overline{z})(w+\overline{w})} , \\
&A_{j}(z) = (z+\overline{z})(x_{j}-(-1)^{j}it
\frac{z-\overline{z}}{4}) + \sum^{\infty}_{k=3}
(z^{k}-(-\overline{z})^{k})c^{(j)}_{k}\ (j = 1,2) , \\
&A_{3}(z,w) = zx_{1} + \overline{w}z_{2} + it (\frac{z^{2}}{4} +
\frac{\overline{w}^{2}}{4}) + \sum^{\infty}_{k=3} (z^{k}c^{(1)}_{k} -
(-\overline{w})^{k} c^{(2)}_{k}) .
\endalign$$
Then $q = -\tau_{1}/\tau_{0}$ and $\varphi = -\frac{1}{2}
(\frac{\partial}{\partial x_{1}} + \frac{\partial}{\partial
x_{2}})^{2} \log \tau_{0}$ is a solution of (4.3.12), where

$$\tau_{1} = a_{3}e^{A_{3}(z_{12},z_{21})} (1 +
a_{1}C(z_{12},z_{11})e^{A_{1}(z_{11})})  (1 +
a_{2}C(\overline{z}_{21},\overline{z}_{22})e^{A_{2}(z_{22})}),
\tag{5.3.5a}$$

$$\aligned
&\tau_{0} = (1+a_{1}e^{A_{1}(z_{11})})
(1+a_{2}e^{A_{2}(z_{22})}) \\
&+
D(z_{12},z_{21})e^{A_{3}(z_{12},z_{21})+\overline{A_{3}(z_{12},z_{21})}}(1 +
a_{1}|C(z_{12},z_{21})|^{2}e^{A_{1}(z_{11})})
\\
&\times (1 + a_{2}|C(z_{21},z_{22})|^{2}e^{A_{2}(z_{22})}).
\endaligned \tag{5.3.5b}$$

Consider now two special cases of (5.3.5a,b):

\vskip 10pt
\noindent
$
(D)\ \ \ z_{1} \equiv z_{11} = z_{12}\ \ \text{and}\ \ z_{2} \equiv
z_{22} =z_{21} ,$
\vskip 10pt
\noindent
$(S)\ \ \ a_{i} = 0\ \ \ \ (i = 1,2),$
\vskip 10pt
\noindent
and let $T = D$ or $S$.  Then (5.3.5a and b) reduce to

$$\tau_{1} = a_{3}e^{A_{3}(z_{1},z_{2})}\ \text{in both cases},
\tag{5.3.6a}$$

$$\tau^{(T)}_{0} = (1 + \delta_{TD}a_{1}e^{A_{1}(z_{1})})(1 +
\delta_{TD}a_{2}e^{A_{2}(z_{2})}) +
D(z_{1},z_{2})e^{A_{1}(z_{1})+A_{2}(z_{2})}, \tag{5.3.6b}$$
so that $q^{(T)} = -\tau_{1}/\tau^{(T)}_{0},\ \varphi^{(T)} =
-\frac{1}{2} (\frac{\partial}{\partial x_{1}} +
\frac{\partial}{\partial x_{2}})^{2} \log \tau^{(T)}_{0}$ is a solution of
(4.3.12).

In order to rewrite $q^{(T)}$ in a more familiar form, let $(j =
1,2\ \text{and}\ a_{j} (z_{j} + \overline{z}_{j}) > 0)$:

$$\aligned
&p^{(T)}_{j} = (a_{j}(z_{j}+\overline{z}_{j}))^{-1/2} \ \text{if}\ T = D\
\text{and}\
=1\ \text{if}\ T = S, \\
&\mu_{1} = \mu_{1R} + i\mu_{1I} = {\tsize{\frac{1}{2}}} \overline{z}_{1},\
\mu_{2} = \mu_{2R} + i\mu_{2I} = {\tsize{\frac{1}{2}}}z_{2}, \\
&m^{(T)}_{j} = 2\sqrt{2}\mu_{jR}p^{(T)}_{j}
e^{-\sum^{\infty}_{k=3}(-1)^{j}((-1)^{j}2\mu_{j})^{k}c^{(j)}_{k} } ,\\
&\xi_{j} = 2x_{j} + 2\mu_{jI}t,\ {\tilde \xi}_{j} =
\frac{1}{\mu_{jR}} \log \frac{|m^{(T)}_{j}|}{\sqrt{2\mu_{jR}}} , \\
&\rho^{(T)} = -a_{3}p^{(T)}_{1}p^{(T)}_{2}.
\endaligned$$
Then we obtain the following expression for $q^{(T)}$:

$$\tfrac{4\rho^{(T)} (\mu_{1R}\mu_{2R})^{1/2} \exp \{
-(\mu_{1R}(\xi_{1}-\tilde{\xi}_{1}) +
\mu_{2R}(\xi_{2}-\tilde{\xi}_{2})) + i(-(\mu_{1I}\xi_{1} +
\mu_{2I}\xi_{2}) + (|\mu_{1}|^{2} + |\mu_{2}|^{2})t + arg\
m_{1}m_{2})\} }
{ ((\delta_{TD} + \exp
(-2\mu_{1R}(\xi_{1}-\tilde{\xi}_{1}))(\delta_{TD} + \exp
(-2\mu_{2R}(\xi_{2}-\tilde{\xi}_{2})) + |\rho^{(T)}|^{2}) } $$
The function $q^{(D)}$ is precisely the $(1,1)$-dromion solution of
the Davey-Stewartson equations (4.3.12) with $\kappa = -1$ found in [FS]
(provided that $\mu_{jR} \in {\Bbb R}_{+}$).  On the other
hand, if we let $\mu_{1I} = \mu_{2I} = 0$, then $q^{(T)}$ reduces to
the $2$-dimensional breather solution found in [BLMP].  Finally,
$q^{(S)}$ is a $1$-soliton solution.

Recall that the dromion solutions of the DS equation were originally
discovered in [BLMP] and [FS] (see also [HH]).  The dromion
solutions of the DS equation were first studied from the point of
view of the spinor formalism by [HMM].

\vskip 10pt
{\bf 5.4.}  Similarly, we obtain the following solutions of the
$2$-dimensional Toda chain (4.2.8):

$$u_{\ell} = \cases
\log (\tau^{(N)}_{\ell + 1}/\tau^{(N)}_{\ell}) &\text{if $-N \leq
\ell \leq N-1$} \\
0 &\text{otherwise}
\endcases \tag{5.4.1}$$
where the $\tau$-functions $\tau^{(N)}_{\ell}$ are obtained from
(5.3.1) by letting all indeterminates $x^{(j)}_{m}$ with $m > 1$ to
be arbitrary constants:

$$\tau^{(N)}_{\ell} = \delta_{\ell ,0} + \sum^{4N}_{r=1}
\sum_{(s_{1},\ldots ,s_{r})} c(s_{1},\ldots ,s_{r})c_{s_{1}}\ldots
c_{s_{r}} \exp \sum^{r}_{k=1} (x_{i_{k}}z_{s_{k}} -
x_{j_{k}}w_{s_{k}} ), \tag{5.4.2}$$
where $(s_{1},\ldots ,s_{r})$ runs over (5.3.2)$_{2}$ and $c_{s}\ (s
\in S)$ are arbitrary constants.

\vskip 10pt
{\bf 5.5.}  Let now $n \geq 3$.  Then we obtain solutions of the
$2+1$ $n$-wave system (4.1.9) as follows.  For $1 \leq i,j \leq n$
let

$$
\aligned
\tau^{(N)}_{ij} &= \delta_{ij} + \sum^{Nn^{2}}_{r=1}
\sum_{(s_{1},\ldots ,s_{r})} c_{s_{1}} \ldots c_{s_{r}} \\
&\ \ \ \ \times \exp \sum^{r}_{k=1} (a_{i_{k}} x + b_{i_{k}} t -
y)z_{s_{k}} - (a_{j_{k}}x + b_{j_{k}}t-y)w_{s_{k}} ,
\endaligned \tag{5.5.1}$$
where $(s_{1},\ldots ,s_{r})$ runs over (5.3.2)$_{n}$ and $c_{s}\ (s
\in S)$ are arbitrary constants.  Then $W_{ij} =
\varepsilon_{ji}\tau_{ij}/\tau_{0}\ (i \neq j)$ is a solution of
(4.1.9), and $Q_{ij} = \varepsilon_{ij}
(a_{i}-a_{j})\tau_{ij}/\tau_{0}\ (i \neq j)$ is a solution of
(4.1.10).

\vskip 10pt
\subheading{\S 6. $m$-reductions of the $n$-component KP hierarchy}

{\bf 6.1.}  Fix a positive integer $m$ and let $\omega = \exp
\frac{2\pi i}{m}$.  Introduce the following $mn^{2}$ fields $(1 \leq
i,j \leq n,\ 1 \leq k \leq m)$ [TV]:

$$\alpha^{(ijk)}(z) \equiv \sum_{p \in {\Bbb Z}} \alpha^{(ijk)}_{p} z^{-p-1}
= :\psi^{+(i)}(z)\psi^{-(j)}(\omega^{k}z): , \tag{6.1.1}$$
where the normal ordering is defined by (2.1.6).  Note that

$$\alpha^{(ijm)}(z) = \alpha^{(ij)}(z), \tag{6.1.2}$$
where $\alpha^{(ij)}(z)$ are the bosonic fields, defined by (2.1.5),
which generate the  affine algebra $g\ell_{n}({\Bbb C})^{\wedge}$ with central
charge $1$ (see (2.1.7)).  It is easy to check that for arbitrary
$m$, the fields $\alpha^{(ijk)}(z)$ generate the affine algebra
$g\ell_{mn}({\Bbb C})^{\wedge}$ with central charge $1$. More
precisely, all the operators $\alpha^{(ijk)}_{p}\ (1 \leq i,j \leq
n,\ 1 \leq k \leq m,\ p \in {\Bbb Z})$ together with $1$ form a basis
of $g\ell_{mn}({\Bbb C})^{\wedge}$ in its representation in $F$ with central
charge $1$, the charge decomposition being the decomposition into
irreducibles.  Hence, using (2.1.14), (2.2.5 and 8), we obtain the
vertex operator realization of this  representation of
$g\ell_{mn}({\Bbb C})^{\wedge}$ in the vector space $B$ (see [TV] for
details).

Now, restricted to the subalgebra $s\ell_{mn}({\Bbb C})^{\wedge}$,
the representation in $F^{(0)}$ is not irreducible any more, since
$s\ell_{mn}({\Bbb C})^{\wedge}$ commutes with all the operators

$$\beta^{(n)}_{k} \overset{\text{def}}\to{=} \frac{1}{n}
\sum^{n}_{j=1} \alpha^{(j,j,m)}_{k},\ \ k \in m{\Bbb Z}.
\tag{6.1.3}$$
In order to describe the irreducible part of the representation of
$s\ell_{mn}({\Bbb C})^{\wedge}$ in $B^{(0)}$ containing the vacuum
$1$, we choose the complementary generators of the oscillator algebra
${\frak a}$ contained in $s\ell_{mn}({\Bbb C})^{\wedge}\ (k \in {\Bbb
Z})$:

$$\beta^{(j)}_{k} = \cases \alpha^{(jjm)}_{k} &\text{if $k \notin
m{\Bbb Z}$,} \\
\frac{1}{j(j+1)} (\alpha^{(11m)}_{k} + \ldots + \alpha^{(jjm)}_{k} -
j\alpha^{(j+1j+1m)}_{k}) &\text{if $k \in m{\Bbb Z}$ and $1 \leq j < n$,}
\endcases \tag{6.1.4}$$
so that the operators (6.1.3 and 4) also satisfy relations (2.1.9).
Then the operators $1,\alpha^{(ijk)}_{p}$ for $i \neq j$, together
with operators (6.1.4) form a basis of $s\ell_{mn}({\Bbb
C})^{\wedge}$.  Hence, introducing the new indeterminates

$$y^{(j)}_{k} = \cases x^{(j)}_{k} &\text{if $k \notin m{\Bbb N}$}, \\
\frac{1}{j(j+1)} (x^{(1)}_{k} + \ldots + x^{(j)}_{k} - jx^{(j+1)}_{k})
&\text{if $k \in m{\Bbb N}$ and $j < n$,} \\
\frac{1}{n} (x^{(1)}_{k} + \ldots + x^{(n)}_{k})
&\text{if $k \in mN$ and $j = n$} ,
\endcases \tag{6.1.5} $$
we have: ${\Bbb C}[x] = {\Bbb C}[y]$ and

$$\sigma (\beta^{(j)}_{k}) = \frac{\partial}{\partial y^{(j)}_{k}}\
\text{and}\ \sigma (\beta^{(j)}_{-k}) = ky^{(j)}_{k} \ \text{if}\ k
> 0. \tag{6.1.6}$$
Now it is clear that the irreducible with respect to
$s\ell_{mn}({\Bbb C})^{\wedge}$ subspace of $B^{(0)}$ containing the
vacuum $1$ is the vector space

$$B^{(0)}_{[m]} = {\Bbb C}[y^{(j)}_{k}|1 \leq j < n,\ k \in {\Bbb
N},\ \text{or}\ j = n,\ k \in {\Bbb N}\backslash m{\Bbb Z}] \otimes
{\Bbb C}[M]. \tag{6.1.7}$$
The vertex operator realization of $s\ell_{mn}({\Bbb C})^{\wedge}$ in
the vector space $B^{(0)}_{[m]}$ is then obtained by expressing the
fields $\alpha^{(ijk)}(z)$ for $i \neq j$ in terms of vertex
operators (2.1.14), which are expressed via the operators (6.1.4), the
operators $Q_{i}Q^{-1}_{j}$ and $\alpha^{(i)}_{0} - \alpha^{(j)}_{0}\
(1 \leq i < j \leq n)$ (see [TV] for details).

The $n$-component KP hierarchy of Hirota bilinear equations on $\tau
\in B^{(0)} = {\Bbb C}[y] \otimes {\Bbb C}[M]$ when restricted to
$\tau \in B^{(0)}_{[m]}$ is called the {\it $m$-th reduced} KP {\it
hierarchy}.  It is obtained from the $n$-component KP hierarchy by
making the change of variables (6.1.5) and putting zero all terms
containing partial derivates by
$y^{(n)}_{m},y^{(n)}_{2m},y^{(n)}_{3m},\ldots$.

It is clear from the definitions and results of \S 3 that the
condition on the $n$-component KP hierarchy to be $m$-th reduced is
equivalent to one of the following equivalent conditions (cf.
[DJKM3]):

$$L(\alpha)^{m}\ \text{is a differential operator}, \tag{6.1.8}$$

$$\sum^{n}_{j=1} \frac{\partial W(\alpha )}{\partial x^{(j)}_{m}} =
z^{m}W(\alpha)  , \tag{6.1.9}$$

$$\sum^{n}_{j=1} \frac{\partial \tau}{\partial x^{(j)}_{m}}
= 0 . \tag{6.1.10}$$
It follows from (6.1.8) that these conditions automatically imply
that all of them hold if $m$ is replaced by any multiple of $m$.

The totality of solutions of the $m$-th reduced KP hierarchy is given
by the following

\proclaim{Proposition 6.1}  Let ${\Cal O}_{[m]}$ be the orbit of $1$
under the (projective) representation of the loop group
$SL_{mn}({\Bbb C}[t,t^{-1}])$ corresponding to the representation of
$s\ell_{mn}({\Bbb C})^{\wedge}$ in $B^{(0)}_{[m]}$.  Then

$${\Cal O}_{[m]} = \sigma ({\Cal O}) \cap B^{(0)}_{[m]}.$$
In other words, the $\tau$-functions of the $m$-th reduced KP hierarchy are
precisely the $\tau$-functions of the KP hierarchy in the variables
$y^{(j)}_{k}$, which are independent of the variables
$y^{(n)}_{m\ell},\ \ell \in {\Bbb N}$.
\endproclaim

\demo{Proof} is the same as of a similar statement in [KP2].\ \ \ $\square$
\enddemo

\vskip 10pt
\noindent
REMARK 6.1.  The above representation of $s\ell_{mn}({\Bbb
C})^{\wedge}$ in $B^{(0)}_{[m]}$ is a vertex operator construction of
the basic representation corresponding to the element of the Weyl
group $S_{mn}$ of $s\ell_{mn}({\Bbb C})$ consisting of $n$ cycles of
length $m$ (see [KP1] and [TV]).  In particular, for $n = 1$ this is
the principal realization [KKLW], and for $m = 1$ this is the
homogeneous realization [FK].  The $m$-th reduced $1$-component KP
was studied in a great detail in [DJKM2] (see also [KP2]).

\vskip 10pt
{\bf 6.2.}  Let $n = 1$.  Then the $2$-reduced KP hierarchy becomes
the celebrated KdV hierarchy on the differential operator
$S \equiv (L^{2})_{+} =  \partial^{2} + u,\ \text{where}\ u =
2u_{1}$:

$$\frac{\partial}{\partial x_{2n+1}} S^{\frac{1}{2}} =
[(S^{n+\frac{1}{2}})_{+} , S^{\frac{1}{2}}],\ n = 1,2,\ldots ,
\tag{6.2.1}$$
the first equation of the hierarchy being the classical
Korteweg-deVries equation

$$4 \frac{\partial u}{\partial t} = \frac{\partial^{3}u}{\partial
x^{3}} + 6u \frac{\partial u}{\partial x}. \tag{6.2.2}$$

Of course, the $3$-reduced KP is the Boussinesq hierarchy, and the
general $m$-reduced KP are the Gelfand-Dickey hierarchies.

\vskip 10pt
{\bf 6.3.}  Let $n = 2$.  The equations of the $1$-reduced
$2$-component KP are independent of $x$, hence equation (4.3.8)
becomes independent of $x$ and $\varphi$ becomes $0$ (see (4.3.9)).
Thus, equation (4.3.8) turns into the decoupled non-linear
Schr\"{o}dinger system (called also the AKNS system):

$$\aligned
i \frac{\partial q}{\partial t} &= -\frac{1}{2}
\frac{\partial^{2}q}{\partial y^{2}} - q^{2}r \\
i \frac{\partial r}{\partial t} &= \frac{1}{2}
\frac{\partial^{2}r}{\partial y^{2}} + qr^{2} .
\endaligned \tag{6.3.1}$$
Thus (6.3.1) is a part of the $1$-reduced $2$-component KP.  For
that reason the $1$-reduced $2$-component KP is sometimes called
the non-linear Schr\"{o}dinger hierarchy.  Of course, under the
constraint (4.3.11), we get the non-linear Schr\"{o}dinger equation

$$i \frac{\partial q}{\partial t} = - \frac{1}{2}
\frac{\partial^{2}q}{\partial y^{2}} - \kappa |q|^{2}q . \tag{6.3.2}$$

Similarly, under the same reduction the $2$-dimensional Toda chain
(4.2.8) turns into the $1$-dimensional Toda chain

$$\frac{\partial^{2}u_{n}}{\partial x^{2}} = e^{u_{n}-u_{n-1}} -
e^{u_{n+1}-u_{n}}\ (\text{here}\ x = 2y^{(1)}_{1}). \tag{6.3.3}$$
Thus, the $1$-dimensional Toda chain is a part of the non-linear
Schr\"{o}dinger hierarchy.  It was studied from the representation
theoretical point of view in [TB].

\vskip 10pt
{\bf 6.4.}  Let $n \geq 3$.  Since the constraint (4.1.4) is contained
among the constraints of the $1$-reduced $n$-component KP
hierarchy, we see that the $1+1$ $n$-wave system (4.1.7) is a part
of the $1$-reduced $n$-component KP hierarchy.  Note also that the
$1$-reduction of the $n$-component KP reduces the $2+1$ $n$-wave
interaction system (4.1.10) into the $1+1$ system (4.1.7).

\vskip 10pt
{\bf 6.5.}  Since the non-linear Schr\"{o}dinger system (6.3.1) is a
part of the $1$-reduced $2$-component KP hierarchy, the $1$-reduced
$n$-component KP hierarchy will be called the $n$-component NLS.
Let us give here its formulation since it is especially simple.

Given a $n \times n$ matrix $C(z) = \sum_{j} C_{j} z^{j}$, we let

$$C(z)_{-} = \sum_{j < 0} C_{j}z^{j},\ C(z)_{+} = \sum_{j \geq 0}
C_{j} z^{j}.$$
Also, given a diagonal complex matrix $a = \text{diag}\ (a_{1},\ldots
,a_{n})$ we let

$$x^{a}_{k} = \sum^{n}_{j=1} a_{k}x^{(j)}_{k},\
\frac{\partial}{\partial x^{a}_{k}} = \sum^{n}_{j=1} a_{k}
\frac{\partial}{\partial x^{(j)}_{k}}.$$
Let ${\frak h}$ denote the set of all traceless diagonal matrices over ${\Bbb
C}$.

The $n$-component NLS hierarchy is the following system on matrix
valued functions
$$P(\alpha ) \equiv P(\alpha ,x,z) = 1 + \sum_{j >0} W^{(j)} (\alpha
,x) z^{-j},\ \alpha \in M,$$
where $x = \{ x^{(a)}_{k}|a \in {\frak h}, \ k =1,2,\ldots\}:$

$$\frac{\partial P(\alpha )}{\partial x^{(a)}_{k}} = -(P(\alpha
)aP(\alpha )^{-1}z^{k})_{-}P(\alpha ) \tag{6.5.1}$$
with additional matching conditions
$$(P(\alpha )R(\alpha - \beta ,z)P(\beta )^{-1})_{-} = 0 ,\ \alpha ,
\beta \in M , \tag{6.5.2}$$
where $R(\gamma ,z) \equiv R^{+}(\gamma ,z)$ is defined by (3.3.10).

This formulation implies the Lax form formulation if we consider
$C^{(a)}(x,z) = P(\alpha )aP(\alpha )^{-1}$ for each $a \in {\frak h}$ and
fixed $\alpha$.  Consider a family of commuting matrix valued
functions of the form

$$C^{(a)} \equiv C^{(a)}(x,z) = a + \sum_{j > 0} C^{(a)}_{j} (x)z^{-j},$$
depending linearly on $a \in {\frak h}$, and let $B^{(a)}_{k} =
(C^{(a)}z^{k})_{+}$.   Then the Lax form of the
$n$-component NLS is

$$\frac{\partial C^{(a)}}{\partial x^{(b)}_{k}} =
[B^{(b)}_{k},C^{(a)} ],\ a,b \in {\frak h},\ k = 1,2,\ldots .
\tag{6.5.3}$$
The equivalent zero curvature form of the $n$-component NLS is

$$\frac{\partial B^{(a)}_{\ell}}{\partial x^{(b)}_{k}} -
\frac{\partial B^{(b)}_{k}}{\partial x^{(a)}_{\ell}} = [B^{(b)}_{k},
B^{(a)}_{\ell}] ,\ a,b \in {\frak h}, \ k,\ell = 1,2,\ldots . \tag{6.5.4}$$

Since for the $1$-reduced $n$-component KP one has: $L = \partial$,
i.e. all $U^{(j)} = 0$, we see from Remark 4.3 that the
$n$-component NLS in the form (6.5.3) contains the following system
of equations on functions $A_{ij} \equiv (C^{E_{jj}}_{1})_{ij}\ (i
\neq j)$:

$$\alignedat2
\frac{\partial A_{ij}}{\partial t_{j}} &= -
\frac{\partial^{2}A_{ij}}{\partial x^{2}_{j}} - 2 \sum_{k \neq j}
A_{ij}A_{jk}A_{kj} , && \ \\
\frac{\partial A_{ij}}{\partial t_{i}} &=
\frac{\partial^{2}A_{ij}}{\partial x^{2}_{i}} + 2 \sum_{k \neq i}
A_{ij} A_{ik} A_{ki} ,  && \ \\
\frac{\partial A_{ij}}{\partial t_{k}} &= A_{ik} \frac{\partial
A_{kj}}{\partial x_{k}} - A_{kj} \frac{\partial A_{ik}}{\partial
x_{k}} &&\ \text{if}\ i \neq k,\ j \neq k, \\
\frac{\partial A_{ij}}{\partial x_{k}} &= A_{ik}A_{kj} &&\text{if}\ i
\neq k,\ j \neq k, \\
\sum_{k} \frac{\partial A_{ij}}{\partial x_{k}} &= \sum_{k}
\frac{\partial A_{ij}}{\partial t_{k}} = 0 . && \
\endalignedat \tag{6.5.5} $$
This reduces to (6.3.1) if $n = 2$.

\vskip 10pt
\noindent
REMARK 6.5. Equations (6.5.1), (6.5.3) and (6.5.4) still make sense if we
consider an arbitrary algebraic group $G$ and a reductive commutative
subalgebra ${\frak h}$ of its Lie algebra ${\frak g}$.  The functions
$P(\alpha )$ take values in $G({\Cal A}((z))$ and the functions
$C^{(a)}$ take values in ${\frak g}({\Cal A}((z)))$.
If $G$ is a simply laced simple Lie group, the element $R(\gamma ,z)
\in G({\Bbb C}[z,z^{-1}])$ in matching conditions (6.5.2) can be
generalized as follows.  Let ${\frak h}$ be a Cartan subalgebra of
${\frak g}$, normalize the Killing form on ${\frak g}$ by the condition that
$(\alpha |\alpha) = 2$ for any root $\alpha$, and identify ${\frak
h}$ with ${\frak h}^{*}$ using this form.  Let $M$ (resp. $L$)
$\subset {\frak h}^{*} = {\frak h}$ be the root (resp. weight)
lattice and let $\varepsilon (\alpha ,\beta):\ M \times M @>>> \{ \pm
1\}$ be a bimultiplicative function such that $\varepsilon (\alpha
,\alpha) = (-1)^{\frac{1}{2} (\alpha |\alpha )}$, $\alpha \in M$.
Define $R(\alpha ,z) \in H({\Bbb C}[z,z^{-1}])$ for each $\alpha$ as
follows:

$$R(\alpha ,z) = c_{\alpha } z^{\alpha}, \tag{6.5.6}$$
where in any finite-dimensional representation $V$ of $G$,
$c_{\alpha} \in H$ and $z^{\alpha} \in H$ for $z \in {\Bbb
C}^{\times}$ are defined by

$$c_{\alpha}(v) = \varepsilon (\beta ,\alpha)v,\ z^{\alpha}(v) =
z^{(\alpha |\beta)}v\ \text{if}\ v \in V_{\beta} . \tag{6.5.7}$$
Note that this GNLS hierarchy is closely related to the Bruhat
decomposition in the loop group $G({\Bbb C}((z)))$.

\vskip 10pt
{\bf 6.6.}  It is clear that we get the $\tau$-function of the
$m$-th reduced $n$-component KP hierarchy if we let in (5.1.9)

$$w_{s} = \omega_{s}z_{s},\ s \in S, \tag{6.6.1}$$
where $\omega_{s}$ are arbitrary $m$-th roots of $1$.  The totality of
$\tau$-functions is (a completion of) the orbit of $1 \in B^{(0)}$
under the group $SL_{mn}({\Bbb C}[t,t^{-1}])$.

\enddocument